\pdfoutput=1
\documentclass[lettersize,journal]{IEEEtran}
\IEEEoverridecommandlockouts
\usepackage{cite}
\usepackage{amsmath,amssymb,amsfonts}
\usepackage{graphicx}
\usepackage{textcomp}
\usepackage{xcolor}
\usepackage{mathtools}
\usepackage{amsmath} 
\usepackage{amssymb}
\usepackage{amsfonts}
\usepackage{verbatim} 
\usepackage{bm,color,soul}
\usepackage{algorithm}
\usepackage{algorithm}
\usepackage{algorithmicx}
\usepackage{algpseudocode}
\usepackage{array}
\usepackage[caption=false,font=normalsize,labelfont=sf,textfont=sf]{subfig}
\usepackage{url}
\usepackage{balance}
\usepackage{hyperref}
\usepackage{mathtools}
\usepackage{amsmath} 
\usepackage{amssymb}
\usepackage{amsfonts}
\usepackage{verbatim} 
\usepackage{bm,color,soul}
\usepackage{cite}
\usepackage{stfloats} 
\usepackage{subeqnarray}
\usepackage{cases}
\usepackage{amsthm}
\usepackage{makecell} 
\usepackage{enumerate}

\newtheorem{lem}{\protect\lemmaname}
\newtheorem{prop}{\protect\propositionname}
\newtheorem{rem}{\protect\remarkname}

\newcommand{\rnum}[1]{\uppercase\expandafter{\romannumeral #1\relax}}
\providecommand{\lemmaname}{Lemma}
\providecommand{\propositionname}{Proposition}
\providecommand{\remarkname}{Remark}
\def\BibTeX{{\rm B\kern-.05em{\sc i\kern-.025em b}\kern-.08em
    T\kern-.1667em\lower.7ex\hbox{E}\kern-.125emX}}
\begin{document}

\title{Energy-Efficient MIMO Integrated Sensing and Communications with On-off Non-transmission Power
}
\author{Guanlin Wu, Yuan Fang, Jie Xu, Zhiyong Feng, and Shuguang Cui
\thanks{G. Wu and J. Xu are with the School of Science and Engineering (SSE) and the Future Network of Intelligence Institute (FNii), The Chinese University of Hong Kong (Shenzhen), Shenzhen 518172, China (e-mail: guanlinwu1@link.cuhk.edu.cn, xujie@cuhk.edu.cn). J. Xu is the corresponding author.}
\thanks{Y. Fang is with the Future Network of Intelligence Institute (FNii), The Chinese University of Hong Kong (Shenzhen), Shenzhen 518172, China (e-mail: fangyuan@cuhk.edu.cn).} 
\thanks{Z. Feng is with School of Information and Communication Engineering, Beijing University of Posts and
Telecommunications, Beijing 100876, China (e-mail: fengzy@bupt.edu.cn).}
\thanks{S. Cui is with the SSE and FNii, The Chinese University of Hong Kong (Shenzhen), Shenzhen 518172, China. He is also affiliated with Peng Cheng Laboratory, Shenzhen, China (e-mail: shuguangcui@cuhk.edu.cn).}}

\maketitle
\begin{abstract}
This paper investigates the energy efficiency of a multiple-input multiple-output (MIMO) integrated sensing and communications (ISAC) system, in which one multi-antenna base station (BS) transmits unified ISAC signals to a multi-antenna communication user (CU) and at the same time use the echo signals to estimate an extended target. We focus on one particular ISAC transmission block and take into account the practical on-off non-transmission power at the BS. Under this setup, we minimize the energy consumption at the BS while ensuring a minimum average data rate requirement for communication and a maximum Cram{\'e}r-Rao bound (CRB) requirement for target estimation, by jointly optimizing the transmit covariance matrix and the ``on'' duration for active transmission. We obtain the optimal solution to the rate-and-CRB-constrained energy minimization problem in a semi-closed form. Interestingly, the obtained optimal solution is shown to unify the spectrum-efficient and energy-efficient communications and sensing designs. In particular, for the special MIMO sensing case with rate constraint inactive, the optimal solution follows the isotropic transmission with shortest ``on'' duration, in which the BS radiates the required sensing energy by using sufficiently high power over the shortest duration. For the general ISAC case, the optimal transmit covariance solution is of full rank and follows the eigenmode transmission based on the communication channel, while the optimal ``on'' duration is determined based on both the rate and CRB constraints. Numerical results show that the proposed ISAC design achieves significantly reduced energy consumption as compared to the benchmark schemes based on isotropic transmission, always-on transmission, and sensing or communications only designs, especially when the rate and CRB constraints become stringent. 
\end{abstract}


\section{Introduction}
Integrated sensing and communications (ISAC) has been recognized as an enabling technology towards future six-generation (6G) wireless networks in both academia and industry, in which wireless communications and radar sensing are integrated into a unified platform for joint design and optimization \cite{9737357,8972666}. In ISAC systems, the transmitters can share the same spectrum resources for both communications and sensing, and they can also jointly design the communication and sensing signals to properly mitigate their mutual interference and even reuse them for the dual purposes \cite{9737357,8972666,9652071}. This thus helps enhance the resource utilization efficiency and improve both sensing and communication performances. In addition, the wireless transceivers can also utilize the sensed information for facilitating the wireless communications design \cite{8999605} and exploit the communication infrastructures to enable advanced networked sensing \cite{9842350}. As a result, by enabling the cooperation and coordination between sensing and communications, ISAC is expected to revolutionize the conventional communication-only wireless networks towards the new one with both communication and sensing functions, thus supporting various emerging Internet of things (IoT) applications \cite{9606831} such as smart home, industrial automation, unmanned aerial vehicles (UAVs) \cite{9916163}, and autonomous vehicles \cite{9127852}.

Among various ISAC design approaches, the exploitation of multiple-input multiple-output (MIMO) for ISAC has attracted growing research interests \cite{9737357}. In particular, the MIMO technique can provide both spatial multiplexing and diversity gains for enhancing the data-rate throughput and transmission reliability for communications \cite{telatar1999capacity}, and it can also provide array and diversity gains to improve the detection and estimation accuracy as well as the spatial resolutions for sensing \cite{li2007mimo,haimovich2007mimo}. In the literature, there have been various research efforts devoted to studying the spectrum-efficient MIMO ISAC designs, with the objective of optimizing the sensing and communication performances by optimizing the transmit waveform and beamforming under given transmit power constraints. (e.g., \cite{9652071,hua2022mimo, 8386661,hua2021optimal,9124713,9531484}). For instance, the authors in \cite{8386661,9124713,hua2021optimal,9531484} studied the MIMO ISAC for simultaneous multiuser communication and target sensing. In \cite{8386661}, the transmit waveforms were optimized to minimize the transmit beampattern matching errors for sensing and minimize the multiuser interference for communications at the same time. In \cite{hua2021optimal} and \cite{9124713}, the joint information and sensing transmit beamforming was optimized to minimize the transmit beampattern matching errors or maximize the transmit beampattern gains for sensing, while ensuring the received signal-to-interference-plus-noise (SINR) constraints at communication users (CUs). Furthermore, the work \cite{9531484} exploited the new rate-splitting multiple access (RSMA) technique for ISAC, in which the beampattern matching errors for sensing and the weighted sum rate for communications were jointly optimized subject to per-antenna transmit power constraints. Besides using transmit beampattern as the sensing performance measure, another line of MIMO ISAC works employed the Cram{\'e}r-Rao bound (CRB) as the sensing performance metric, which serves as a lower bound of variance for any unbiased estimators and provides the fundamental performance limits for target estimation. For instance, the authors in \cite{9652071} studied the transmit beamforming design towards joint radar sensing and multi-user communications, in which the CRB  for target estimation was optimized, while ensuring the minimum SINR requirements for individual CUs. Furthermore, the authors in \cite{hua2022mimo} studied the MIMO ISAC with one multi-antenna CU and one target to be estimated, in which the transmit covariance was optimized to reveal the fundamental tradeoff between the data rate for communications and the CRB for sensing, by characterizing the Pareto boundary of the rate-CRB region.

The ever-increasing sensing and communication requirements, however, result in significant network energy consumption and carbon emissions in wireless networks. As such, how to meet the sensing and communication quality of service (QoS) requirements in an energy- and carbon-efficient manner is becoming another important task for 6G ISAC networks \cite{8922617}. Towards this end, the investigation of the energy-efficient and green ISAC is of great significance in practice. Different from the spectrum-efficient ISAC designs \cite{9652071,8386661,hua2021optimal,hua2022mimo,9124713,9531484} that only considered the transmit power, the on-off non-transmission power is an important factor that should be taken into account in the design of energy-efficient ISAC. In particular, by considering a base station (BS) transmitter in practice, the non-transmission power comes from radio frequency (RF) chains, analog-to-digital converters (ADCs), digital-to-analog converters (DACs), etc. If the BS is active in transmission, then these components need to be turned on, which consumes the non-transmission power; while if the BS is inactive without transmission, then these components can be turned off for saving the non-transmission power. It has been shown in energy-efficient communications \cite{6082514,5683485,6409501,6514948} that in order to deliver data bits over a wireless channel most energy efficiently, the BS transmitter should first transmit with an optimized covariance that maximizes the  bits-per-Joule energy efficiency (defined as the ratio of the data rate to the total energy consumption), and then turn off the components for saving the non-transmission power. To our best knowledge, however, how to implement MIMO sensing and MIMO ISAC in an energy-efficient manner has not been well investigated in the literature yet. 

It is worth noting that there have been a handful of prior works \cite{kaushik2021hardware,kaushik2022green,he2022energy,9797869} studying the energy-efficient ISAC under different setups, in which the bits-per-Joule energy efficiency for communications was maximized subject to the beampattern constraints for radar sensing \cite{kaushik2021hardware,kaushik2022green,he2022energy,9797869}. However, these prior works \cite{kaushik2021hardware,kaushik2022green,he2022energy,9797869} assumed the non-transmission power as a constant term for optimization, but did not consider its on-off control over time. Furthermore, these works adopted the transmit beampattern gains as the objective for sensing performance optimization, which cannot directly reflect the target estimation performance. To fill in such a research gap in this work, we are motivated to investigate the energy-efficient MIMO sensing and MIMO ISAC by considering the on-off control of non-transmission power for energy saving, and employing the estimation CRB as the fundamental sensing performance metric. 

In particular, this paper studies an energy-efficient MIMO ISAC system, in which one multi-antenna BS sends unified ISAC signals to communicate with a multi-antenna CU and simultaneously uses the echo signals to estimate an extended target. We focus our study on a particular ISAC transmission block corresponding to the refreshing time for sensing in practice, over which the BS aims to estimate the complete target response matrix. By taking into account the on-off non-transmission power at the BS, we investigate the energy efficiency of this MIMO ISAC system. The main results are listed as follows. 
\begin{itemize}
  \item Our objective is to minimize the total energy consumption at the BS over the whole ISAC block, while ensuring a minimum average data rate constraint for communication and a maximum CRB constraint for target estimation. Towards this end, we jointly optimize the transmit covariance matrix at the BS and the ``on'' duration for its active transmission.
  \item First, we consider the specific case with MIMO sensing only, in which the rate constraint for communications becomes negligible. As the estimation CRB is a convex trace inverse function with respect to the transmit covariance, it is shown that the optimal energy-efficient sensing solution is to employ the isotropic transmission (with identical covariance matrix and proper transmit power) together with the shortest ``on'' duration to minimize the non-transmission energy consumption.
  \item Next, we consider the general ISAC case with both sensing and communications. In this case, we leverage the Lagrange duality method to obtain the optimal solution to the rate-and-CRB-constrained energy minimization problem in a semi-closed form. It is revealed that the optimal transmit covariance is of full rank and follows the eigenmode transmission based on the communication channel. Furthermore, proper power allocations are employed over different eigenmodes, and the "on" duration is optimized depending on the rate and CRB constraints. Interestingly, the obtained optimal solution is observed to unify the energy- and spectrum-efficient communication and sensing designs. 
  \item Finally, we present numerical results to validate the performance of our proposed optimal ISAC solution, as compared to the benchmark schemes based on isotropic transmission, always-on transmission, as well as the sensing or communications only designs. It is shown that the proposed optimal solution significantly outperforms other benchmark schemes, especially when the communications and sensing requirements become stringent. It is also shown that the isotropic transmission design performs close to the optimal solution, thus showing its effectiveness in practical implementation.
\end{itemize}

\textit{Notations}: Vectors and matrices are denoted by bold lower- and upper-case letters, respectively. $\boldsymbol{I}$ represents an identity matrix with proper dimensions. For a complex arbitrary-size matrix $\boldsymbol{A}$, $\text{rank}(\boldsymbol{A})$, $\boldsymbol{A}^T$, and $\boldsymbol{A}^H$ denote its rank, transpose, and conjugate transpose, respectively. For a square matrix $\boldsymbol{Q}$, $\text{tr}(\boldsymbol{Q})$, $\text{det}(\boldsymbol{Q})$, and $\text{rank}(\boldsymbol{Q})$ represent its trace, determinant, and rank, respectively, and $\boldsymbol{Q} \succeq \boldsymbol{0}$ means that $\boldsymbol{Q}$ is positive semi-definite. $\mathbb{E}(\cdot )$ denotes the statistical expectation. $\otimes$ denotes the Kronecker product. $\circ$ denotes the Hadamard product.

\section{System Model and Problem Formulation}
\subsection{Signal Model}
\begin{figure*}[t]
  \centering
  \includegraphics[width=123mm]{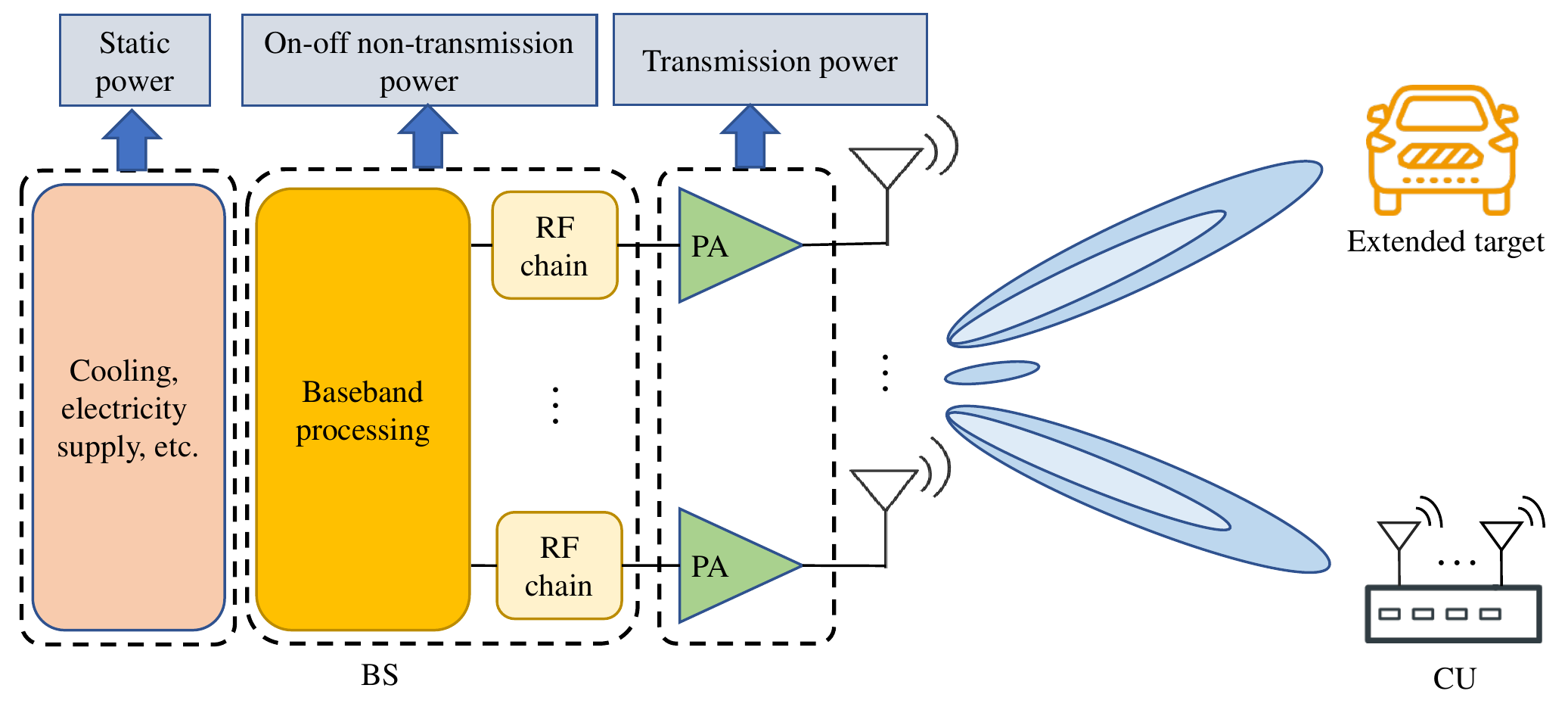}
  \caption{Illustration of the MIMO ISAC system and the power consumption model at BS.}
  \label{system model}
\end{figure*}
We consider a MIMO ISAC system as shown in Fig.~\ref{system model}, in which a BS sends unified ISAC signals to a CU and estimates an extended target based on its echo signal. Suppose that the BS is equipped with $M \geq 1$ transmit antennas for sending ISAC signals and $N_s\geq M$ receive antennas for sensing. The CU is equipped with $N_c\geq 1$ receive antennas. 

In particular, we focus on the ISAC transmission over a particular finite block with duration $T_{\text{max}}$, during which the wireless channels remain unchanged. Here, $T_{\text{max}}$ is assumed to be sufficiently large for facilitating the ISAC design. We consider the ``on-off'' transmission at the BS for enhancing the energy efficiency. Suppose that the BS is active in transmission over the ``on'' period of duration $\tau$, and keeps silent to save energy in the remaining time. Here, $\tau$ is a continuous optimization variable to be determined later, which satisfies $T_{\text{min}} \le \tau \le T_{\text{max}}$. Note that $T_{\text{min}}$ denotes the minimum duration for active transmission, which is set for meeting the ISAC requirements. Furthermore, for notational convenience, let $B$ denote the signal bandwidth and $1/B$ denote the symbol duration. Accordingly, we denote $\hat{\tau}=\tau B $ as the number of transmitted symbols over this block, and denote $\hat{T}_{\text{min}}=T_{\text{min}}B$ and $\hat{T}_{\text{max}}=T_{\text{max}}B$ as the minimum number of symbols that can be transmitted and the total number of symbols within the whole block, respectively. 

Let $\boldsymbol{x}(t) \in \mathbb{C}^{M \times 1}$ denote the transmit ISAC signal at the BS over the $t$-th symbol, $t\in\{1,\ldots,\hat{\tau}\}$. It is assumed that $\boldsymbol{x}(t)$ is a circularly symmetric complex Gaussian (CSCG) random vector with zero mean and covariance matrix $\boldsymbol{Q}=\mathbb{E}(\boldsymbol{x}(t)\boldsymbol{x}^H(t))$. This assumption is made in order for achieving the capacity of MIMO communication. Furthermore, let $\boldsymbol{X}=[\boldsymbol{x}(1),\ldots,\boldsymbol{x}(\hat{\tau})] $ denote the accumulated ISAC signals over this block in matrix form.
 
First, we consider the MIMO communications from the BS to the CU. Let $\boldsymbol{H}_c \in \mathbb{C}^{N_c\times M}$  denote the MIMO channel matrix from the BS to the CU, with rank($\boldsymbol{H_c})=r\leq M$. The received signal by the CU at symbol $ t\in \{1,\ldots,\hat{\tau} \}$ is
\begin{align}
    \boldsymbol{y}_c(t)=\boldsymbol{H}_c\boldsymbol{x}(t)+\boldsymbol{z}_c(t),
\end{align}
where $\boldsymbol{z}_c(t)$ denotes the noise at the CU receiver that is a CSCG random vector with zero mean and covariance $\sigma_c^2\boldsymbol{I}$. Under the Gaussian signalling with transmit covariance $\boldsymbol{Q}$ at the BS, the average achievable communication rate (in bits/second/Hertz, bps/Hz) at the CU over this block is 
\begin{align}
  R(\boldsymbol{Q})=\frac{\tau}{T_{\text{max}}}\text{log}_2\text{det}\left(\boldsymbol{I}_{N_c}+\frac{\boldsymbol{H_c}\boldsymbol{Q}\boldsymbol{H_c}^H}{\sigma_c^2}\right).
\end{align}
In order to facilitate the transmit design, it is assumed that the BS transmitter and the CU receiver perfectly know the channel state information (CSI) of $\boldsymbol{H}_c$ via proper channel estimation and feedback. 

Next, we consider the MIMO sensing. In particular, we consider the sensing estimation of an extended target, in which the target is regarded as an object with multiple scatters. Accordingly, the target response matrix  $\boldsymbol{H_s}$ from the BS transmitter to the target to the BS receiver is given as
\begin{align}
  \boldsymbol{H}_s=\sum_{k=1}^K\zeta_k\boldsymbol{b}(\phi_k)\boldsymbol{a}^H(\theta_k),
\end{align}
where $K$ denotes the number of scatterers, $\zeta_k$ denotes the reflection coefficient of the $k$-th scatterer that depends on its the radar cross section (RCS), and $\boldsymbol{a}(\theta_k)\in \mathbb{C}^{M \times 1}$ and $\boldsymbol{b}(\phi_k) \in \mathbb{C}^{N_s \times 1}$ denote the transmit and receive steering vectors with angle of departure (AoD) $\theta_k$ and angle of arrival (AoA) $\phi_k$, respectively. During the active transmission period, the received radar signal $\boldsymbol{Y}_s \in \mathbb{C}^{N_s\times \hat{\tau}}$ at the sensing receiver of BS is
\begin{align}
    \boldsymbol{Y}_s=\boldsymbol{H}_s \boldsymbol{X}+\boldsymbol{Z}_s,
\end{align}
where $\boldsymbol{Z}_s \in \mathbb{C}^{N_s\times \hat{\tau}}$ denotes the noise matrix at the BS, each element of which is a CSCG random variable with zero mean and variance $\sigma^2_s$. In this work, we focus on estimating the complete target response matrix $\boldsymbol{H}_s$, based on which we can further extract the angle and reflection coefficients of different scatterers via algorithms like the multiple signal classification ($\text{MUSIC}$) \cite{li1996adaptive,schmidt1986multiple}. We consider the CRB as the performance metric to characterize the fundamental sensing performance for target estimation, which serves as the variance lower bound of any practical biased estimators \cite{kay1993fundamentals}. As shown in \cite{9652071}, the CRB matrix for estimating $\boldsymbol{H}_s$ is given by
\begin{align}\label{CRB matrix for Hs}
  \text{CRB}(\boldsymbol{H_s})=\left(\frac{1}{\sigma_s^2}(\boldsymbol{X}^T)^H\boldsymbol{X}^T\otimes \boldsymbol{I}_{N_s}\right)^{-1}.
\end{align}
Notice that by assuming the minimum sensing duration $T_{\text{min}}$ and accordingly the active ISAC duration $\hat{\tau}$ to be sufficiently large, the sample covariance matrix can be safely approximated as the statistical covariance matrix \cite{4276989}, i.e., we have \vspace{-0.5cm}
\begin{align}\label{approximation Q}
\boldsymbol{Q} \approx \frac{\boldsymbol{X}\boldsymbol{X}^H}{\hat{\tau}} = \frac{\boldsymbol{X}\boldsymbol{X}^H}{B\tau }.
\end{align}
As such, the CRB matrix is further expressed as 
\begin{align} \label{CRB fur}
  \text{CRB}(\boldsymbol{H_s})=\left(\frac{B\tau }{\sigma_s^2}\boldsymbol{Q}^T\otimes  \boldsymbol{I}_{N_s}\right)^{-1}.
\end{align}
Based on \eqref{CRB fur},  we use the trace of the CRB matrix as the sensing performance metric, which characterizes the sum CRB for estimating the elements of $\boldsymbol{H}_s$ \cite{4838872}, i.e.,
\begin{align}\label{trace of CRB}
  \text{tr}(\text{CRB}(\boldsymbol{H_s}))=\frac{\sigma_s^2N_s}{B \tau}\text{tr}(\boldsymbol{Q}^{-1}).
\end{align}
\subsection{Power Consumption Model}
Next, we consider the practical power consumption model at the BS for facilitating the energy-efficient  MIMO ISAC design. The power consumption at the BS is generally divided into three parts, namely the transmission power, on-off non-transmission power, and static power, respectively.
\subsubsection{Transmission power}
The transmission power generally comes from the power amplifiers (PAs) for signal radiation. Let $0< \eta < 1$ denote the drain efficiency of the PA. Recall that $\text{tr}(\boldsymbol{Q})$ denote the transmit power by the BS. Then we have the transmission power as $P_{\text{trans}}=\frac{1}{\eta }\text{tr}(\boldsymbol{Q})$. 
\subsubsection{On-off non-transmission power}
The non-transmission power generally comes from the non-ideal circuits in the RF chains, ADCs/DACs, filters, and baseband signal processing components. The non-transmission power has the following on-off property. If the BS transmitter is turned on with $\text{tr}(\boldsymbol{Q})> 0$, then these components need to be activated  for consuming the non-transmission power; while if the BS transmitter is switched off with $\text{tr}(\boldsymbol{Q}) = 0$, then these components can be turned off for saving power. Let $P_c$ denote a constant power consumed during the ``on'' status. We have the on-off non-transmission power as 
\begin{align}
  {P_{\text{non-trans}}} = \left\{ \begin{array}{l}
    P_c,~~\text{if}~\text{tr}(\boldsymbol{Q}) > 0,\\
    0,~~~~\text{if}~\text{tr}(\boldsymbol{Q}) = 0.
    \end{array} \right.
\end{align}
\subsubsection{Static power}
The static power $P_{\text{static}}$ is a constant term that is consumed by, e.g., cooling systems and electricity supply at the BS.

By combining the above three parts and considering the on duration $\tau$, the total energy consumption by the BS over the whole ISAC block is given by  
\begin{align}\label{total energy consumption}
  P_{\text{total}}(\boldsymbol{Q}, \tau) =  \left(\frac{\text{tr}(\boldsymbol{Q})}{\eta}+P_c\right) \tau + P_{\text{static}}T_{\text{max}}.
\end{align}
\subsection{Problem Formulation}
Our objective is to minimize the total energy consumption $P_{\text{total}}(\boldsymbol{Q}, \tau)$ at the BS in \eqref{total energy consumption} while ensuring the minimum rate requirement $R$ for communications and the maximum CRB threshold $\Gamma$ for sensing, by optimizing both the transmit covariance $\boldsymbol{Q}$ and the ``on'' transmission duration $\tau$ for active transmission. As a result, the rate-and-CRB-constrained energy minimization problem for MIMO ISAC is mathematically formulated as 
\begin{subequations}
 \begin{align}
  &(\text{P1}):&{\mathop \text{min}\limits_{\boldsymbol{Q}\succeq  0, \tau}}\quad&{\left(\frac{\text{tr}(\boldsymbol{Q})}{\eta}+P_c\right) \tau}\label{Original Problem1}\\
  &&{\text{s.t.}} \quad&{\frac{\sigma_s^2N_s}{ B\tau}\text{tr}(\boldsymbol{Q}^{-1})\leq \Gamma \quad \quad} \label{CRB constraint P1}\\
  &&&{\frac{ \tau}{T_{\text{max}}}\text{log}_2\text{det}\left(\boldsymbol{I}_{N_c}+\frac{1}{\sigma_c^2}\boldsymbol{H_c}\boldsymbol{Q}\boldsymbol{H_c}^H\right)}\geq R \label{Rate constraint P1}\\
  &&&  T_{\text{min}} \leq \tau \leq T_{\text{max}}.\quad\quad \label{Q and tau constraint P1}
\end{align} 
\end{subequations}
Notice that in problem (P1), we remove the constant static energy consumption $P_{\text{static}}T_{\text{max}}$ in \eqref{total energy consumption} without loss of optimality. 

\section{Optimal Solution to Problem (P1) in Special Communications or Sensing Dominated cases} \label{Sec special}
Before presenting the optimal solution to problem (P1), in this section we consider two special cases when the communications and sensing are dominated, in which the CRB constraint in \eqref{CRB constraint P1} for sensing and the rate constraint in \eqref{Rate constraint P1} for communications become inactive, respectively. They also correspond to that there is only communication or sensing in this system, respectively. 
\subsection{Communications-dominated Case} \label{Sec ComD}
First, we consider the communications-dominated case, in which the CRB constraint in \eqref{CRB constraint P1} can be skipped. In this case, problem (P1) is reduced as
\begin{subequations}
\begin{align}\label{only communication} 
  (\text{P2}):~&{\mathop \text{min}\limits_{\boldsymbol{Q}\succeq  0, \tau}}\quad{\left(\frac{\text{tr}(\boldsymbol{Q})}{\eta}+P_c\right) \tau}\\
    &~~~{\text{s.t.}} ~\quad{\frac{\tau}{T_{\text{max}}}\text{log}_2\text{det}\left(\boldsymbol{I}_{N_c}+\frac{1}{\sigma_c^2}\boldsymbol{H_c}\boldsymbol{Q}\boldsymbol{H_c}^H\right)}\geq R \label{rate constraint for only Com} \\
    &\quad\quad\quad~~ \eqref{Q and tau constraint P1} \nonumber.
\end{align}
\end{subequations} 
Notice that at the optimality of problem (P2), the rate constraint \eqref{rate constraint for only Com} must be met with equality, i.e., we have $\tau = R T_{\text{max}}/\text{log}_2\text{det}\left(\boldsymbol{I}_{N_c}+\frac{1}{\sigma_c^2}\boldsymbol{H_c}\boldsymbol{Q}\boldsymbol{H_c}^H\right)$. By substituting this, problem (P2) is equivalent to 
\begin{subequations}
\begin{align}\label{EE maximization}
  (\text{P2.1}):~&\mathop {{\rm{max}}}\limits_{{\boldsymbol{Q}\succeq  0}} \quad \frac{{\text{log}_2\text{det}\left(\boldsymbol{I}_{N_c}+\frac{1}{\sigma_c^2}\boldsymbol{H_c}\boldsymbol{Q}\boldsymbol{H_c}^H\right)}}{RT_{\text{max}}({\frac{1}{\eta}\text{tr}(\boldsymbol{Q})+P_c})}\\
  &~~{\text{s.t.}}~~R \leq {\text{log}_2\text{det}\left(\boldsymbol{I}_{N_c}+\frac{\boldsymbol{H_c}\boldsymbol{Q}\boldsymbol{H_c}^H}{\sigma_c^2}\right)} \leq \frac{RT_{\text{max}}}{T_{\text{min}}},  \label{on duration constraint to UC}
\end{align}
\end{subequations}
which corresponds to a rate-constrained  bits-per-Joule  energy efficiency maximization problem that has been investigated in \cite{5683485}. To solve problem (P2.1), we need to consider the following two problems, namely the unconstrained bits-per-Joule energy efficiency maximization and the rate-constrained transmit power minimization problems, respectively. 
\begin{align}\label{EE maximization unconstrained}
    (\text{P2.2}):~&\mathop \text{max}\limits_{{\boldsymbol{Q}\succeq  0}} \quad \frac{{\text{log}_2\text{det}\left(\boldsymbol{I}_{N_c}+\frac{1}{\sigma_c^2}\boldsymbol{H_c}\boldsymbol{Q}\boldsymbol{H_c}^H\right)}}{RT_{\text{max}}({\frac{1}{\eta}\text{tr}(\boldsymbol{Q})+P_c})}.
\end{align}
\begin{subequations}
    \begin{align}\label{transmit power minimization}
      (\text{P2.3}):~&{\mathop \text{min}\limits_{\boldsymbol{Q}\succeq  0}}\quad{\frac{\text{tr}(\boldsymbol{Q})}{\eta}+P_c}\\
      &~{\text{s.t.}} \quad {\text{log}_2\text{det}\left(\boldsymbol{I}_{N_c}+\frac{\boldsymbol{H_c}\boldsymbol{Q}\boldsymbol{H_c}^H}{\sigma_c^2}\right)} \geq \bar{R},
    \end{align}
\end{subequations}
where $\bar{R} $ corresponds to the minimum rate constraint that is a parameter in problem (P2.3). Let $\boldsymbol{Q}_{\text{EE-com}}$ denote the optimal solution to problem (P2.2), and $\boldsymbol{Q}_{\text{SE-com}}(\bar{R})$ denote the optimal solution to problems (P2.3) with rate constraint $\bar{R}$. Furthermore, we denote the singular value decomposition (SVD) of communication channel $\boldsymbol{H}_c$ as
\begin{align} \label{SVD channel}
  \boldsymbol{H}_c=\boldsymbol{U}_c\boldsymbol{\Sigma} \boldsymbol{V}_c^H,
\end{align}
where $\boldsymbol{\Sigma}=\text{diag}(\lambda_1,\ldots,\lambda_r,0,\ldots,0) \in \mathbb{R}^{N_c \times M}$ with $\lambda_1 \geq \lambda_2 \geq \ldots \geq \lambda_r >0$ denoting the $r$ positive singular values, and $\boldsymbol{U}_c\in \mathbb{C}^{N_c \times N_c}$ and $\boldsymbol{V}_c\in \mathbb{C}^{M \times M}$ with $\boldsymbol{U}_c^H\boldsymbol{U}_c=\boldsymbol{U}_c\boldsymbol{U}_c^H=\boldsymbol{I}_{N_c}$ and $\boldsymbol{V}_c^H\boldsymbol{V}_c=\boldsymbol{V}_c\boldsymbol{V}_c^H=\boldsymbol{I}_{M}$. Then we have the following lemmas from prior work \cite{5683485}, for which the proofs are omitted here. 
\begin{lem} \label{unconstrained solutions}
\textup{The optimal solution $\boldsymbol{Q}_{\text{EE-com}}$ to problem (P2.2) is given as
\begin{align}
  \boldsymbol{Q}_{\text{EE-com}}={\boldsymbol{V}_c}{{\boldsymbol{S}}_{\text{EE-com}}}{{\boldsymbol{V}}_c^H},
\end{align}
where ${{\boldsymbol{S}}_{\text{EE-com}}}$ is a diagonal matrix with diagonal elements ${\left[ {{\boldsymbol{S}}_{\text{EE-com}}} \right]_{ii}}$ given in
\begin{align} \label{diagonal elements SEE}
  {\left[ {{\boldsymbol{S}}_{\text{EE-com}}} \right]_{ii}} = \left\{ \begin{array}{l}
    \frac{\eta }{{\xi^* RT_{\text{max}}\text{ln} 2}} - \frac{{\sigma _c^2}}{{\lambda _i^2}},~i = 1,...,r\\
    0,\quad\quad\quad\quad~ i = r+1,...,M,
    \end{array} \right.
\end{align}
$\xi^*$ represents the auxiliary parameter that can be uniquely derived based on the following equation via bisection search. 
\begin{align}
  \log _2\text{det}\left( {{{\boldsymbol{I}}_{{N_c}}} + \frac{1}{{\sigma _c^2}}{{\boldsymbol{H}}_{c}}{{\boldsymbol{Q}}_{\text{EE-com}}}{{\boldsymbol{H}}_{c}}^H} \right)  \nonumber\\ 
  -\xi^* RT_{\text{max}}\left( {\frac{1}{\eta}\text{tr}({{\boldsymbol{Q}}_{\text{EE-com}}}) + {P_c}} \right) = 0.
\end{align}
}
\end{lem}
\begin{lem} \label{power minimization solutions}
\textup{The optimal solution $\boldsymbol{Q}_{\text{SE-com}}(\bar{R})$ to problem (P2.3) is
\begin{align} 
  \boldsymbol{Q}_{\text{SE-com}}(\bar{R})={\boldsymbol{V}_c}{{\boldsymbol{S}}_{\text{SE-com}}(\bar{R})}{{\boldsymbol{V}}_c^H},
\end{align}
where $\boldsymbol{S}_{\text{SE-com}}(\bar{R})$ is a diagonal matrix with diagonal elements ${\left[ \boldsymbol{S}_{\text{SE-com}}(\bar{R}) \right]_{ii}}$ given by 
\begin{align} \label{diagonal elements SSE}
  {\left[ {{\boldsymbol{S}}_{\text{SE-com}}(\bar{R})} \right]_{ii}} = \left\{ \begin{array}{l}
    \frac{\eta }{{q^*(\bar{R}) \text{ln} 2}} - \frac{{\sigma _c^2}}{{\lambda _i^2}},~i = 1,...,r\\
    0,\quad\quad\quad\quad~ i = r+1,...,M,
    \end{array} \right.
\end{align}
and $q^*(\bar{R})$ is the solution to the equality
\begin{align}
  \text{log} _2\text{det}\left(\boldsymbol{I}_{N_c}+\frac{1}{\sigma_c^2} \boldsymbol{H}_c \boldsymbol{Q}_{\text{SE-com}}(\bar{R}) \boldsymbol{H}_c{ }^H\right)=\bar{R}.
\end{align}
}
\end{lem}
Notice that in Lemma \ref{unconstrained solutions}, $\boldsymbol{Q}_{\text{EE-com}}$ corresponds to the energy-efficient communications design, which maximizes the bits-per-Joule energy efficiency or minimizes the energy consumption for delivering each unit bit. With $\boldsymbol{Q}_{\text{EE-com}}$, we denote the correspondingly achieved communication rate as 
\begin{align}
  \varUpsilon (\boldsymbol{Q}_{\text{EE-com}})=\text{log}_2\text{det}\left(\boldsymbol{I}_{N_c}+\frac{\boldsymbol{H_c}\boldsymbol{Q}_{\text{EE-com}}\boldsymbol{H_c}^H}{\sigma_c^2}\right),
\end{align}
which will be used for solving problem (P2) later. By contrast, in Lemma \ref{power minimization solutions}, $\boldsymbol{Q}_{\text{SE-com}}(\bar{R})$ can be viewed as the spectrum-efficient communications design, since this design efficiently utilizes the spectrum resources for meeting the rate constraint with the minimum transmit power. Based on Lemmas \ref{unconstrained solutions} and \ref{power minimization solutions}, we have the following proposition for solving problem (P2). 
\begin{prop} \label{Optimal solution only communication case}
\textup{The optimal solution $\boldsymbol{Q}^*_{\text{com}}$ and $\tau^*_{\text{com}}$ to problem (P2) in the communications-dominated case is given as follows by considering three different cases.
\begin{itemize}
  \item If $ \varUpsilon (\boldsymbol{Q}_{\text{EE-com}})>R \frac{T_{\text{max}}}{T_{\text{min}}} $, we have \begin{align}
      {\boldsymbol{Q}}_{\textup{com}}^* = \boldsymbol{Q}_{\text{SE-com}}\left(\frac{RT_{\text{max}}}{T_{\text{min}}}\right), ~~\tau_{\text{com}}^*=T_{\text{min}}.
  \end{align}
  \item If $ R\leq\varUpsilon (\boldsymbol{Q}_{\text{EE-com}}) \leq R\frac{T_{\text{max}}}{T_{\text{min}}}$, we have 
  \begin{align}
    \boldsymbol{Q}_{\text{com}}^*=\boldsymbol{Q}_{\text{EE-com}}, \quad \tau_{\text{com}}^*=\frac{RT_{\text{max}}}{\varUpsilon (\boldsymbol{Q}_{\text{EE-com}})}.
  \end{align}
  \item If $\varUpsilon (\boldsymbol{Q}_{\text{EE-com}})<R$, we have
  \begin{align}
    {{\boldsymbol{Q}}_{\textup{com}}^*} = \boldsymbol{Q}_{\text{SE-com}}(R), ~~\tau_{\text{com}}^*=T_{\text{max}}.
  \end{align}
\end{itemize}
}
\begin{proof}
This proposition can be easily verified by noting the fact that the energy efficiency objective in \eqref{EE maximization} is monotonically increasing with respect to the rate $\text{log}_2\text{det}\left(\boldsymbol{I}_{N_c}+\frac{\boldsymbol{H_c}\boldsymbol{Q}\boldsymbol{H_c}^H}{\sigma_c^2}\right)$ when it is less than $\varUpsilon (\boldsymbol{Q}_{\text{EE-com}})$, and monotonically decreasing when the rate is greater than $\varUpsilon (\boldsymbol{Q}_{\text{EE-com}})$. Therefore, the details are omitted for brevity. 
\end{proof}
\end{prop}
It is revealed from Proposition \ref{Optimal solution only communication case} that the optimal solution to problem (P2) in the communications-dominated case unifies  the energy-efficient and spectrum-efficient communications designs. When the rate requirement $R$ is sufficiently large with $R > \varUpsilon (\boldsymbol{Q}_{\text{EE-com}})$ or sufficiently small with $\frac{T_{\text{max}}}{T_{\text{min}}}R<\varUpsilon (\boldsymbol{Q}_{\text{EE-com}}) $, the BS needs to be turned on over the whole block by setting $\tau_{\text{com}}^* = T_{\text{max}}$ or over the shortest ``on'' duration by setting $\tau_{\text{com}}^* = T_{\text{min}}$, respectively, in which the spectrum-efficient communications design is implemented to meet the rate requirement. By contrast, when $R\leq\varUpsilon (\boldsymbol{Q}_{\text{EE-com}}) \leq R\frac{T_{\text{max}}}{T_{\text{min}}}$, the energy-efficient communications design is desired, and the optimized on duration $\tau_{\text{com}}^*$ is between $T_{\text{min}}$ and $T_{\text{max}}$, such that the tradeoff between the transmission versus non-transmission energy consumption is properly balanced. Finally, it is worth noting that for the special case with $P_c=0$, we have $\varUpsilon (\boldsymbol{Q}_{\text{EE-com}}) = 0$ and accordingly $\tau^*_{\text{com}}=T_{\text{max}}$. In general, when other parameters are given, the value of the optimal ``on'' duration $\tau_{\text{com}}^*$ is monotonically non-decreasing with respect to the non-transmission power $P_c$. 
\begin{rem} \label{com dominate}
  \textup{Notice that at the optimal solution to problem (P2) with communications only, the resultant $\text{CRB}$ is $\text{CRB}_{\text{com}}=\frac{\sigma_s^2N_s}{ B\tau_{\text{com}}^*}\text{tr}(\boldsymbol{Q}_{\text{com}}^{*-1})$. If $\text{CRB}_{\text{com}} $ is no greater than $\Gamma$ in (P1), then $Q^*_{\text{com}}$ and $\tau^*_{\text{com}}$ are actually the optimal solution to problem (P1). This corresponds to a trivial case of (P1), in which the communication constraint dominates the sensing requirement. Nevertheless, if  $\text{CRB}_{\text{com}} > \Gamma$, then the obtained solution of $\boldsymbol{Q}_{\text{com}}^*$ and $\tau_{\text{com}}^*$ are not feasible for (P1). For instance, if $\boldsymbol{Q}_{\text{com}}^*$ is rank deficient, then $\text{CRB}_{\text{com}}$ becomes infinite, and thus $\text{CRB}_{\text{com}} > \Gamma$ always holds. We will address this non-trivial case in Section \ref{Optimal section}.}
\end{rem} 
\subsection{Sensing-dominated Case}
Next, we consider the sensing-dominated case, in which the rate constraint in problem (P1) is inactive. In this case, we  minimize the total energy consumption under the maximum CRB constraint. Accordingly, by dropping the rate constraint \eqref{Rate constraint P1} in (P1), the sensing-constrained energy minimization problem becomes 
\begin{subequations}
\begin{align}\label{only sensing} 
    (\text{P3}):~&{\mathop \text{min}\limits_{\boldsymbol{Q}\succeq  0, \tau}}\quad{\left(\frac{\text{tr}(\boldsymbol{Q})}{\eta}+P_c\right) \tau}\\
    &\quad {\text{s.t.}} ~\quad{\frac{\sigma_s^2N_s}{ B\tau}\text{tr}(\boldsymbol{Q}^{-1})\leq \Gamma \quad \quad}\label{sensing performance constraint only sensing}\\
    &\quad\quad\quad~~  \eqref{Q and tau constraint P1}\nonumber.
\end{align}
\end{subequations}
We have the following proposition for solving problem (P3). 
\begin{prop} \label{Optimal solution only sen}
\textup{The optimal solution to (P3) is given by }
\begin{align}
  {{\boldsymbol{Q}}_{\textup{sen}}^*}= \frac{\sigma_s^2N_sM}{BT_{\textup{min}}\Gamma}\boldsymbol{I}_M,~~{\tau_{\textup{sen}}^*} = T_{\textup{min}}. \label{SensingD opt}
\end{align}
\begin{proof}
 See Appendix \ref{Appendix only sen}.
\end{proof}
\end{prop}
It is observed from Proposition \ref{Optimal solution only sen} that the energy-efficient MIMO sensing is achieved by transmitting with the shortest ``on'' duration. This can be intuitively explained as follows. Note that the CRB function in \eqref{trace of CRB} is inversely proportional to the transmission energy over the block. As a result, by minimizing the ``on'' duration and increasing transmission power, the BS transmitter can meet the CRB requirement with the same transmission energy but minimized non-transmission energy. This phenomenon is different from that for the communications-dominated case in Section \ref{Sec ComD}, in which longer ``on'' duration is preferred when the rate constraint $R$ becomes large, as the communication rate is actually a concave function with respect to the transmission energy.

\begin{rem} \label{sensing dominate}
  \textup{At the optimal solution to problem (P3) with sensing only, the resultant rate is $R_{\text{sen}}=\frac{T_{\text{min}}}{T_{\text{max}}}\text{log}_2\text{det}\left( {{{\boldsymbol{I}}_{{N_c}}} + \frac{\sigma_s^2N_sM}{BT_{\textup{min}}\Gamma}{{\boldsymbol{H}}_{c}}{{\boldsymbol{H}}_{c}}^H} \right)$. If $R_{\text{sen}}$ is no smaller than the rate constraint $R$ in problem (P1), then $\boldsymbol{Q}^*_{\text{sen}}$ and $\tau_{\text{sen}}^*$ are also optimal for problem (P1). This corresponds to another trivial case of (P1), where the sensing constraint dominates the communications requirement.}
\end{rem}
\section{Optimal Solution to Problem (\text{P1}) in General ISAC Case} \label{Optimal section}
This section presents the optimal solution to the rate-and-CRB constrained energy minimization problem (P1). Notice that for the communications-dominated case with $\Gamma \ge \text{CRB}_{\text{com}}$ and the sensing-dominated case with $R\leq R_{\text{sen}}$, the optimal solutions to problem (P1) have been obtained in Proposition \ref{Optimal solution only communication case} and \ref{Optimal solution only sen}, respectively. In this section, we focus on the general case when both sensing and communications constraints are active. 
\subsection{Problem Reformulation}
First, we transform (P1) into a convex optimization problem. Towards this end, we introduce $\boldsymbol{E}=\boldsymbol{Q} \tau$. Problem (P1) is thus reformulated as
\begin{subequations}
  \begin{align} 
      &(\text{P4}):&{\mathop \text{min}\limits_{\boldsymbol{E}\succeq  0, \tau}}\quad&{\frac{\text{tr}(\boldsymbol{E})}{\eta}+P_c\tau } \\
      &&{\text{s.t.}} \quad&{\frac{\sigma_s^2N_s}{B}\text{tr}(\boldsymbol{E}^{-1})\leq \Gamma \quad }  \label{crb reformulated}\\ 
      &&&{\frac{\tau}{T_{\text{max}}}\text{log}_2\text{det}\left(\boldsymbol{I}_{N_c}+\frac{\boldsymbol{H}_c \boldsymbol{E}\boldsymbol{H}_c^H}{\sigma_c^2 \tau}\right)}\geq R \label{rate reformulated}\\ 
      &&&  \eqref{Q and tau constraint P1}. \nonumber
  \end{align}
\end{subequations}
In problem (P4), constraint \eqref{crb reformulated} is convex, constraint \eqref{Q and tau constraint P1} is linear, and constraint \eqref{rate reformulated} is convex as the left-hand side is a concave perspective function. Therefore, problem (P4) is a convex problem that can be solved by standard convex optimization techniques. To gain more insights, in the following, we derive its optimal solution in a well-structured form. Based on the SVD of the communication channel $\boldsymbol{H}_c$ in \eqref{SVD channel}, we define 
\begin{align}
    \tilde{\boldsymbol{Q}}&=\boldsymbol{V}_c^H\boldsymbol{Q}\boldsymbol{V}_c ,\label{transform for Q}\\
    \tilde{\boldsymbol{E}}&=\tilde{\boldsymbol{Q}} \tau =\boldsymbol{V}_c^H\boldsymbol{E}\boldsymbol{V}_c.\label{transform for E}
\end{align}
Accordingly, problem (P4) is equivalently reformulated as
\begin{subequations}
\begin{align} 
    &(\text{P4.1}):&{\mathop \text{min}\limits_{\tilde{\boldsymbol{E}}\succeq  0, \tau}}\quad&{\frac{\text{tr}(\tilde{\boldsymbol{E}})}{\eta}+P_c\tau } \label{P2 transformation for objective function}\\
    &&{\text{s.t.}} \quad&{\frac{\sigma_s^2N_s}{B}\text{tr}(\tilde{\boldsymbol{E}}^{-1})\leq \Gamma \quad } \label{P2 transformation for CRB} \\
    &&&{\frac{\tau}{T_{\text{max}}}\text{log}_2\text{det}\left(\boldsymbol{I}_{N_c}+\frac{\boldsymbol{\Sigma}^2\tilde{\boldsymbol{E}}}{\sigma_c^2 \tau}\right)}\geq R \label{P2 transformation for rate}\\
    &&&  \eqref{Q and tau constraint P1}. \nonumber
\end{align}
\end{subequations}
In problem (P4.1), \eqref{P2 transformation for objective function} and \eqref{P2 transformation for CRB} are obtained based on the fact that $\text{tr}(\boldsymbol{E})=\text{tr}(\boldsymbol{V}_c\tilde{\boldsymbol{E}}\boldsymbol{V}_c^H)=\text{tr}(\boldsymbol{V}_c^H\boldsymbol{V}_c\tilde{\boldsymbol{E}})=\text{tr}(\tilde{\boldsymbol{E}})$ and $\text{tr}(\boldsymbol{E}^{-1})=\text{tr}((\boldsymbol{V}_c\tilde{\boldsymbol{E}}\boldsymbol{V}_c^H)^{-1})={\text{tr}}({{\boldsymbol{V}}_c}{\tilde {\boldsymbol{E}}^{ - 1}}{\boldsymbol{V}}_c^H) = {\text{tr}}({\boldsymbol{V}}_c^H{{\boldsymbol{V}}_c}{\tilde {\boldsymbol{E}}^{ - 1}}) = {\text{tr}}({\tilde {\boldsymbol{E}}^{ - 1}})$. In addition, \eqref{P2 transformation for rate} is obtained based on $\text{det}(\boldsymbol{I}_{N_c}+\frac{1}{\sigma_c^2}\boldsymbol{H_c}\boldsymbol{Q}\boldsymbol{H_c}^H)=\text{det}(\boldsymbol{I}_{N_c}+\frac{1}{\sigma_c^2 \tau}\boldsymbol{H_c}\boldsymbol{E}\boldsymbol{H_c}^H)=\text{det}(\boldsymbol{I}_{N_c}+\frac{1}{\sigma_c^2 \tau}\boldsymbol{U}_c\boldsymbol{\Sigma } \boldsymbol{V}_c^H\boldsymbol{E}\boldsymbol{V}_c\boldsymbol{\Sigma}^{H}\boldsymbol{U}^{H}_c )=\text{det}(\boldsymbol{I}_{N_c}+\frac{\boldsymbol{\Sigma}^2\tilde{\boldsymbol{E}}}{\sigma_c^2 \tau})$. Next, we have the following lemma.
\begin{lem} \label{lemma diagonal}
\textup{The optimal solution $\tilde{\boldsymbol{E}}$ to problem (P4.1) is a diagonal matrix with all diagonal elements being positive, i.e., $\tilde{\boldsymbol{E}}=\text{diag}(\boldsymbol{e})=\text{diag}(e_1,e_2,\ldots,e_M)$ with $e_i>0,\forall i\in\{1,\ldots,M\}$ denoting the diagonal elements.
}
\begin{proof}
\textup{See Appendix \ref{Appendix diagonal}.
} 
\end{proof}
\end{lem}
 
Based on Lemma \ref{lemma diagonal}, problem ($\text{P4.1}$) is further rewritten as the following power allocation problem:
\begin{subequations}
  \begin{align}
    &(\text{P4.2}):&{\mathop \text{min}\limits_{\left\{e_i \geq 0\right\}, \tau}}\quad&{\frac{1}{\eta }\sum\limits_{i = 1}^M {{e_i}}  + {P_c}\tau}\\
    &&{\text{s.t.}} ~~\quad&{\frac{\sigma _s^2{N_s}}{B}\sum\limits_{i = 1}^M {\frac{1}{{{e_i}}}}  \le {\Gamma}} \label{crb constraint}\\
    &&&\frac{\tau }{T_{\text{max}}}\sum\limits_{i = 1}^r {{\rm{lo}}{{\rm{g}}_2}\left( {1 + \frac{{\lambda _i}^2{e_i}}{{\sigma _c^2\tau }}} \right)}  \ge R \label{perspective function convex}\\
    &&& \eqref{Q and tau constraint P1} .\nonumber
  \end{align}
\end{subequations}
\subsection{Optimal Power Allocation Solution to Problem (P4.2)}
We apply the Lagrange duality method to derive the well-structured optimal power allocation solution to problem (P4.2) for gaining more insights. Let $\gamma\geq0$ and $\nu\geq0$ denote the dual variables associated with the constraints in \eqref{crb constraint} and \eqref{perspective function convex}, respectively. The partial Lagrangian of (P4.2) is 
\begin{align}\label{Lagrangian for primal problem}
    &\mathcal{L} (\boldsymbol{e},  \tau, \gamma, \nu)=\frac{1}{\eta }\sum\limits_{i = 1}^M {{e_i}}  + \gamma \left( \frac{\sigma _s^2{N_s}}{B}\sum\limits_{i = 1}^M {\frac{1}{{{e_i}}}}  - {\Gamma}   \right) \nonumber\\
    &+ {P_c}\tau -\nu \left( \frac{\tau }{T_{\text{max}}}\sum\limits_{i = 1}^r {{\rm{lo}}{{\rm{g}}_2}\left( {1 + \frac{{\lambda _i}^2{e_i}}{{\sigma _c^2\tau }}} \right)}  - {R}\right).
\end{align}
The corresponding dual function is given by 
  \begin{align}\label{dual function of primal problem}
    \mathcal{G} (\gamma, \nu)={\mathop \text{min}\limits_{\left\{e_i \geq 0\right\}, \tau}} \mathcal{L} (\boldsymbol{e}, \tau, \gamma, \nu)\quad \text{s.t.}~\eqref{Q and tau constraint P1}.
  \end{align}
Therefore, the dual problem of ($\text{P4.2}$) is given by 
    \begin{align} \label{dual problem of primal problem}
     (\text{D4.2}):\quad \mathop \text{max}\limits_{ \gamma\geq0, \nu\geq0} &\quad\mathcal{G}( \gamma, \nu).
   \end{align} 
Denote $(\gamma^*, \nu^* )$ as the optimal solution to dual problem (D4.2). Since problem (P4.2) is convex and satisfies the Slater's condition, it satisfies the strong duality and can be solved by equivalently solving its dual problem. By solving the dual problem (D4.2), we have the following proposition.
\begin{prop}\label{Proposition of relation between ei and t}
\textup{The optimal solution to problem (P4.2) is given by $(e_i^*,\tau^*)$ in the following.  
\begin{itemize}
  \item For the last $M-r$ subchannels, i.e., subchannels $ i\in\{r+1,\ldots, M\}$, we have
  \begin{align} \label{ei r1}
      e^*_{i}=\sqrt{\frac{\eta{\gamma^*}\sigma _s^2{N_s}}{B}}, ~\forall i\in \{r+1,\ldots,M\}.
  \end{align}
  \item For the first $r$ subchannels, i.e., $ i \in \{1,\ldots,r\}$, $e_i^*$ and $\tau^*$ satisfy
  \begin{align}
    \frac{1}{\eta}-\frac{\gamma^*\sigma_s^2N_s}{B}\frac{1}{(e_i^*)^2}-\frac{\nu^*}{\text{ln}2T_{\text{max}}}\left(\frac{\frac{\lambda_i^2}{\sigma_c^2}}{1+\frac{\lambda_i^2e_i^*}{\sigma_c^2\tau^*}}\right)=0. \label{power within r}
  \end{align}
  Accordingly, $e_i^*$ can be expressed as a function of $\tau^*$, denoted by
  \begin{align}
    e_i^*= \hat{e}_i(\tau^*)&=-k_1+\sqrt[3]{-k_2+\sqrt{k_2^2+k_3^3}} \nonumber \\
    &+\sqrt[3]{-k_2-\sqrt{k_2^2+k_3^3}}, \quad \forall i\in \{1,\ldots,r\}, \label{cubic equation}
\end{align} 
where $k_1=\frac{b_i}{3 a_i} \tau^*, k_2=\frac{27 a_i^2 d \tau^*-9 a_i b_i c_i \tau^*+2\left(b_i \tau^*\right)^3}{54 a_i^3}$, and $k_3= \frac{3a_ic_i-(b_i\tau^*)^2}{9a_i^2}$, with $a_i= \frac{\lambda_i^2}{\eta\sigma_c^2}$, $b_i=\frac{1}{\eta}-\frac{\nu^*}{\text{ln}2T_{\text{max}}} \frac{\lambda_i^2}{\sigma_c^2}$, $c_i=d \frac{\lambda_i^2}{\sigma_s^2}$, and $d=-\frac{\gamma^* \sigma_s^2 N_s}{B}$.
 \item Furthermore, the optimal ``on'' duration $\tau^*$ is given by
 \begin{align}\label{tau^*}
\tau^* = \text{min}(\text{max}(\bar{\tau}, T_{\text{min}}), T_{\text{max}}),
\end{align}
where $\bar{\tau}$ denotes the solution of the following equality: 
 \begin{align} 
  &P_c-\frac{\nu^*}{T_{\text{max}}}\left(\sum_{i=1}^r\textup{log}_2(1+\frac{\lambda_i^2 \hat{e}_i(\tau)}{\sigma_c^2\tau})\right. \nonumber \\
  & \quad\quad\quad \left. +\frac{1}{\textup{ln}2}\sum_{i=1}^r(\frac{1}{1+\frac{\lambda_i^2\hat{e}_i(\tau)}{\sigma_c^2\tau}})-\frac{r}{\textup{ln}2}\right)=0. \label{bisection equation} 
\end{align}
\end{itemize}
}
\begin{proof}
  See Appendix \ref{Appendix Lagrange}.
\end{proof}
\end{prop} 
Based on Proposition \ref{Proposition of relation between ei and t}, we can first obtain $e_i^*$ under given $\tau$ based on \eqref{ei r1} and \eqref{cubic equation}, and then find the optimal $\tau^*$ based on \eqref{bisection equation} via bisection, as the left-hand side of \eqref{bisection equation} is monotonic with respect to $\tau$. Therefore, the optimal solutions $\left\{e_i^*\right\}$ and $\tau^*$ to problem (P4.2) are obtained.

With $\left\{e_i^*\right\}$ and $\tau^*$ at hands, accordingly we obtain $\tilde{\boldsymbol{E}}^*=\textup{diag}(e_1^*,\ldots,e_M^* )$. Based on \eqref{transform for Q} and \eqref{transform for E}, the optimal $\boldsymbol{Q}^*$ of problem ($\text{P1}$) is finally derived as
\begin{align} \label{finally transmit covariance}
  \boldsymbol{Q}^*= \frac{\boldsymbol{V}_c\tilde{\boldsymbol{E}}^*\boldsymbol{V}_c^{H}}{\tau^*}.
\end{align}

The obtained optimal transmit covariance solution $\boldsymbol{Q}^*$ provides interesting insights. First, based on \eqref{finally transmit covariance}, we express $\boldsymbol{Q}^* = \boldsymbol{V}_c\tilde{\boldsymbol{Q}}^*\boldsymbol{V}_c^H$, where $\tilde{\boldsymbol{Q}}^*=\frac{\tilde{\boldsymbol{E}}^*}{\tau^*} =\textup{diag}(p_1^*,\ldots,p_M^*)$, with $p_i^*=\frac{e_i^*}{\tau^*}$ representing the power allocation. This shows that the optimal transmit covariance matrix follows the eigenmode transmission based on the eigenmodes of communication channel $\boldsymbol{V}_c$, where the first $r$ eigenmodes or subchannels are used for both sensing and communications, and the last $M-r$ eigenmodes or subchannels are used for dedicated sensing only. Next, it is observed from \eqref{ei r1} and \eqref{cubic equation} that the power allocations $\left\{e_i^*\right\}$ or $\left\{p_i^*\right\}$ over the first $r$ subchannels depend on the corresponding channel gains $\left\{\lambda_i\right\}$, and equal power allocation is adopted over the last $M-r$ subchannels. In particular, we have the following proposition.
\begin{prop} \label{Prop power allocation}
\textup{The optimal $\left\{e_i^*\right\}$, $i\in \left\{1,\ldots,M\right\}$ satisfies $e_1^*\geq e_2^*\geq \ldots e_r^*\geq e_{r+1}^*= \ldots =e_M^* >0$. Equivalently, the optimal transmit power allocation $\left\{p_i^*\right\}$ satisfies $p_1^*\geq p_2^*\geq \ldots p_r^*\geq p_{r+1}^*= \ldots =p_M^* >0$. 
\begin{proof}
  See Appendix \ref{Appendix power allocation}.
\end{proof}
}
\end{prop}

Next, it is worth noting that the proposed energy-efficient ISAC design unifies the energy- and spectrum-efficient communications and sensing designs. In particular, the optimal solution to (P1) in Proposition \ref{Proposition of relation between ei and t} contains the energy-efficient communications design $\boldsymbol{Q}_{\text{EE-com}}$ and the spectrum-efficient communications design $\boldsymbol{Q}_{\text{SE-com}}$ as special cases. In particular, when  $\Gamma > \text{CRB}_{\text{com}}$, i.e., the CRB constraint \eqref{crb constraint} is inactive, the corresponding dual variable $\gamma^*=0$. Then from \eqref{power within r} we obtain the water-filling power allocation that is same as the optimal solution in Proposition \ref{Optimal solution only communication case} for the communications-dominated case. Similarly, the optimal solution in Proposition \ref{Proposition of relation between ei and t} also contains the energy-efficient sensing design $\boldsymbol{Q}_{\text{sen}}$ as another special case. In particular, when $ R < R_{\text{sen}}$, i.e., the rate constraint in \eqref{perspective function convex} is inactive, the corresponding dual variable $\nu^*=0$. It is observed from \eqref{cubic equation} and \eqref{bisection equation} that power allocations on different subchannels are identical and $\tau^*= T_{\text{min}}$ holds, such that the ISAC case is reduced to the sensing-dominated case as showed in Proposition \ref{Optimal solution only sen}.  

Furthermore, it is interesting to discuss the special case with $P_c = 0$, i.e., only transmission power is considered. In order to minimize the transmission energy consumption in this case, the ``on'' duration must be as long as possible and thus we have $\tau^*=T_{\text{max}}$ based on \eqref{bisection equation} and \eqref{tau^*}. Accordingly, the always-on transmission is optimal for (P1), together with proper transmit covariance design. In addition, we discuss another special case when the rate constraint $R$ is sufficiently large and communication channel $\boldsymbol{H}_c$ is of full rank, 
(i.e., $r=M$). In this case, the optimal transmit covariance $\boldsymbol{Q}_{\text{com}}^*= \boldsymbol{Q}_{\text{SE-com}}(R)$ in the communications-dominated case will be reduced to isotropic transmission with equal power allocation over subchannels, as $q^*(R)$ in \eqref{diagonal elements SSE} becomes sufficiently small. As such, $\boldsymbol{Q}_{\text{com}}^*$ in the communications-dominated case coincides with $\boldsymbol{Q}_{\text{sen}}^*$ in the sensing-dominated case. Therefore, it follows that $\boldsymbol{Q}^*$ for the ISAC case also becomes the isotropic transmission in this case. 

\section{Numerical Results}
This section presents numerical results to show the performance of our proposed ISAC designs under different setups. For comparison, we also consider the following benchmark schemes.
\begin{itemize}
  \item \textbf{Isotropic transmission}: We set the transmit covariance as $\boldsymbol{Q}=p\boldsymbol{I}_M$, with $p$ representing the transmit power, that is an optimization variable to be decided. By replacing $\boldsymbol{Q}$ with $p\boldsymbol{I}_M$ in problem (P1), we optimize both $p$ and $\tau$ to minimize the total energy consumption while ensuring the rate and CRB constraints. 
  \item \textbf{Communication-based design}: This scheme is motivated by the water-filling-like power allocation in Proposition \ref{Optimal solution only communication case} for the communications-dominated case. This design is applicable only when channel matrix $\boldsymbol{H}_c$ is of full rank, such that the corresponding communications-dominated transmit covariance $\boldsymbol{Q}_{\text{com}}^*$ is also of full rank in order to ensure the CRB requirement for sensing. In particular, based on the communications-dominated design with $\boldsymbol{Q}_{\text{com}}^*$ and $\tau_{\text{com}}^*$, we set $\boldsymbol{Q}= \alpha\boldsymbol{Q}_{\text{com}}^*$  and $\tau = \tau_{\text{com}}^*$, in which $\alpha$ is a scaling factor to meet the CRB constraint, i.e.,  
  \begin{align}
    \alpha=\left(\frac{\sigma _s^2{N_s}}{B \tau^{\text{com}}}\text{tr}((\boldsymbol{Q}_{\text{com}}^*)^{-1}) \right)/\Gamma.
  \end{align}
  \item \textbf{Sensing-based design}: This scheme is motivated by the optimal solution of $\boldsymbol{Q}_{\text{sen}}^*=\frac{\sigma_s^2N_sM}{BT_{\textup{min}}\Gamma}\boldsymbol{I}_M$ and $\tau_{\text{sen}}^* =T_{\text{min}}$ for the sensing-dominated case. Accordingly, the BS sets $\tau=T_{\text{min}}$ and $\boldsymbol{Q}=p\boldsymbol{I}_M$, in which $p$ is the transmit power to be determined to meet the CRB and rate constraints. Specifically, we have
  \begin{align}
    p = \text{max}(p_1^{\text{sen}},p_2^{\text{sen}}),
  \end{align}
  where $p_1^{\text{sen}}$ and $p_2^{\text{sen}}$ are the solutions to equalities $\frac{T_{\text{min}} }{T_{\text{max}}}\sum\limits_{i = 1}^r {{\rm{lo}}{{\rm{g}}_2}\left( {1 + \frac{{\lambda _i}^2{p_1^{\text{sen}}}}{{\sigma _c^2}}} \right)}=R$ and $\frac{\sigma_s^2N_sM}{ B T_{\text{min}}p_2^{\text{sen}}}= \Gamma$, respectively.
  \item \textbf{Always-on transmission}: The BS is turned ``on'' for active transmission over the whole ISAC block, i.e., $\tau=T_{\text{max}}$. As such, we only need to design the transmit covariance $\boldsymbol{Q}$ for minimizing total energy consumption, similarly as that in Section \ref{Optimal section}. 
\end{itemize}

In the simulation, we set the number of transmit antennas at the BS as $M=6$ and that of the received antenna for sensing at the BS as $N_s=8$. We also set the signal bandwidth as $B=10~\text{MHz}$, the noise power as $\sigma^2_c=\sigma^2_s=-103 \text{dBm}$, the PA efficiency as $\eta=0.38$, and the non-transmission power as $P_c=45 \text{W}$. We consider Rician fading for the communication channel $\boldsymbol{H}_c$ with the Rician factor equal to 1, for which the pathloss follows $51.2+41.2\text{log}_{10}d$ with $d$ (in meters) denoting the distance between the BS and the CU. We set the minimum number of symbols for transmission as $\hat{T}_{\text{min}}=150$ and the total number of symbols over this block as $\hat{T}_{\text{max}}=256$. Therefore, the minimum active transmission duration is $T_{\text{min}}=\hat{T}_{\text{min}}/B=15 \mu \text{s}$, and the total block duration is $T_{\text{max}}=\hat{T}_{\text{max}}/B=25.6 \mu\text{s}$, which are reasonable based on the practical coherent processing interval (CPI) in radar sensing \cite{richards2014fundamentals}.
\subsection{Case with Full-rank Communication Channel}  \label{Sec full}
\begin{figure}[h]
  \centering
  \includegraphics[width=80mm]{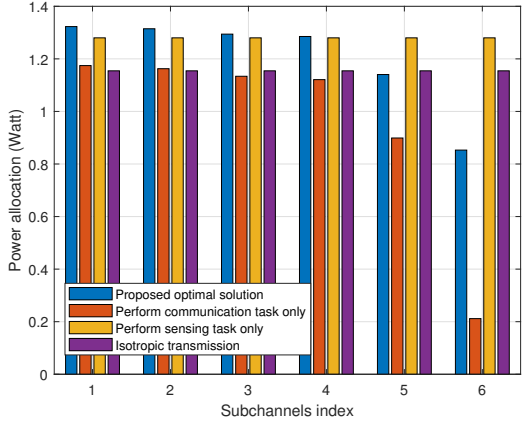}
  \caption{Power allocation behaviors with $M=N_c=6$, $\Gamma=0.25$, and $R=18\text{bps/Hz}$.}
  \label{allocation full}
\end{figure}
\begin{figure}[h]
  \centering
  \includegraphics[width=80mm]{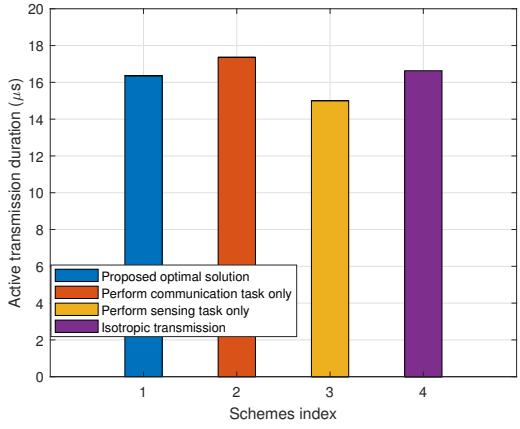}
  \caption{Active transmission duration by four schemes with $M=N_c=6$, $\Gamma=0.25$, and $R=18\text{bps/Hz}$.}
  \label{durationfullrank}
\end{figure}
\begin{figure}[h]
  \centering
  \includegraphics[width=80mm]{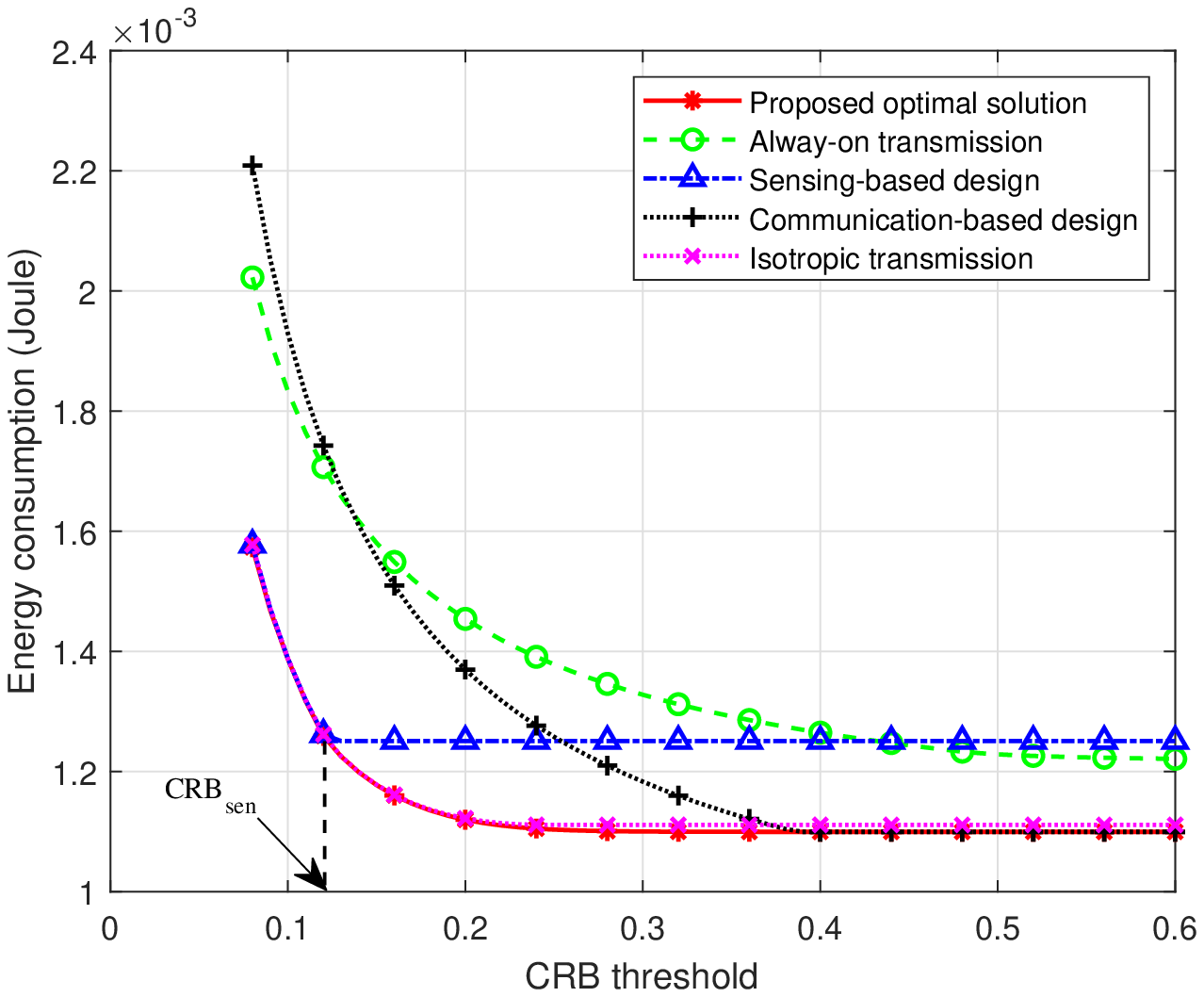}
  \caption{Energy consumption versus CRB threshold $\Gamma$ with $R=18\text{bps/Hz}$ and $d=100\text{m}$.}
  \label{CRBvery}
\end{figure}
\begin{figure}[h]
  \centering
  \includegraphics[width=80mm]{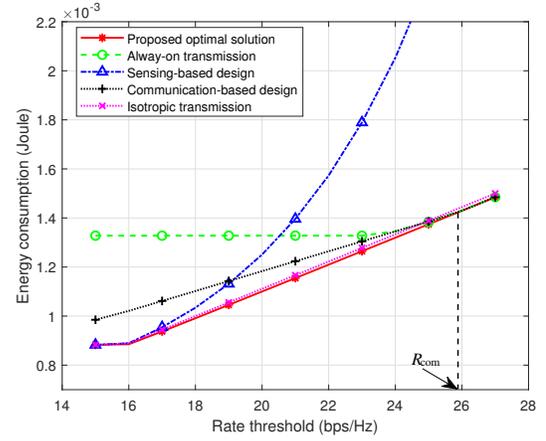}
  \caption{Energy consumption versus rate threshold $R$ with $\Gamma=0.3$ and $d=100\text{m}$.}
  \label{ratevery}
\end{figure}
First, we consider the case with full-rank communications channel $\boldsymbol{H}_c$, in which the number of receive antennas at CU is set as $N_c=M=6$. 

Fig.~\ref{allocation full} shows the power allocation behaviors achieved by the proposed optimal solution and the heuristic isotropic transmission as compared to the schemes when the sensing and communications are only performed, where $\Gamma=0.25$ and $R=18\text{bps/Hz}$. Furthermore, Fig.~\ref{durationfullrank} shows the correspondingly obtained optimal ``on'' duration for active transmission under these schemes. It is observed from Fig.~\ref{allocation full} that for the proposed optimal solution, the achieved power allocations are monotonically non-increasing over the $r=M=6$ subchannels, which is consistent with Proposition \ref{Prop power allocation}. It is also observed in Fig.~\ref{allocation full} that the transmit power at the first four subchannels by the proposed optimal solution is larger than those by the communication task only and sensing task only, and in Fig.~\ref{durationfullrank} that the ``on'' duration by proposed optimal solution is between those by the communication task only and sensing task only. This shows the effectiveness of our proposed solution in unifying both designs for meeting the CRB and rate constraints. For the isotropic transmission, it is observed in Fig.~\ref{allocation full} that the resultant power allocations over the first four subchannels are lower than those under the proposed optimal solution, but those over the last two subchannels are higher. Furthermore, its resultant ``on'' duration in Fig.~\ref{durationfullrank} is observed to be slightly higher than that under the proposed optimal solution, due to its suboptimality in optimization. 

Fig.~\ref{CRBvery} shows the energy consumption versus the CRB threshold $\Gamma$ under the given rate constraint $R=18\text{bps/Hz}$. It is observed that the energy consumption achieved by the proposed optimal solution is significantly less than those by the other benchmark schemes. It is also observed that the performance by the isotropic transmission performs close to the proposed optimal solution, especially when the value of $\Gamma$ is small. This shows the effectiveness of this design in this case. In addition, when $\Gamma$ becomes sufficiently small (e.g., $\Gamma \leq \text{CRB}_{\text{sen}}$), the proposed optimal solution is observed to perform same as that by the sensing-based design. This is due to the fact that the resultant rate $R_{\text{sen}}$ is no smaller than the rate constraints $R=18\text{bps/Hz}$, such that the sensing constraint dominates the communication one in this case, and the proposed optimal solution is reduced to the sensing-based design (see Remark \ref{sensing dominate}). By contrast, when $\Gamma$ is high, the communication-based design is observed to achieve similar performance as the proposed optimal solution, as communication is dominant in this case.

Fig.~\ref{ratevery} shows the energy consumption versus the rate threshold $R$ under the given CRB constraint $\Gamma=0.3$. It is observed that the energy consumption achieved by the proposed optimal solution is less than those by the other benchmark schemes. It is worth noting that when $R$ becomes sufficiently large, the always-on transmission performs similarly as that by communication-based design. This is due to the fact that $R$ is sufficiently large in this case, such that the whole ISAC block should be utilized for transmission, and the transmit covariance optimization becomes identical for the two schemes. Furthermore, it is also observed that when $R$ is greater than $R_{\text{com}}$, the proposed optimal solution is observed to achieve the same energy consumption as that by the communication-based design. This is due to the fact that the achieved $\text{CRB}_{\text{com}}$ is no greater than the CRB constraint $\Gamma=0.3$, such that the communication rate constraint dominates the sensing one in this case, and the proposed optimal solution is reduced to the communication-based design (see Remark \ref{com dominate}). 
\subsection{Case with Rank-deficient Communication Channel}
\begin{figure}[ht]
  \centering
  \includegraphics[width=80mm]{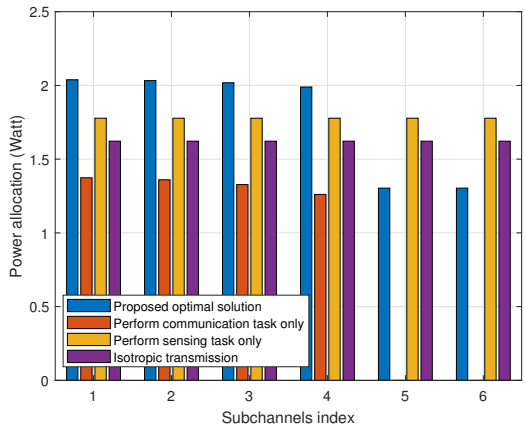}
  \caption{Power allocation behaviors with $M=6, N_c=4$, $\Gamma=0.18$, and $R=15\text{bps/Hz}$.}
  \label{allocation}
\end{figure}
\begin{figure}[ht]
  \centering
  \includegraphics[width=80mm]{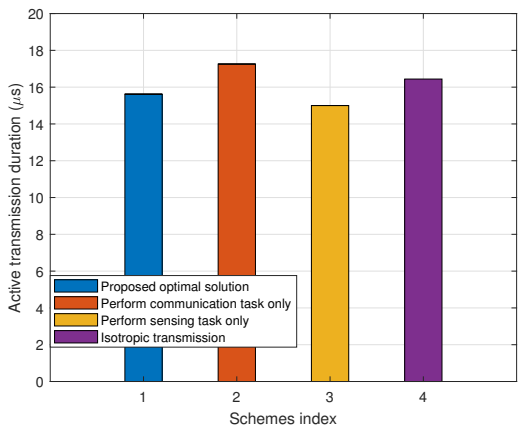}
  \caption{Active transmission duration by four schemes with $M=6, N_c=4$, $\Gamma=0.18$, and $R=15\text{bps/Hz}$.}
  \label{durationrankdeficient}
\end{figure}
\begin{figure}[ht]
  \centering
  \includegraphics[width=80mm]{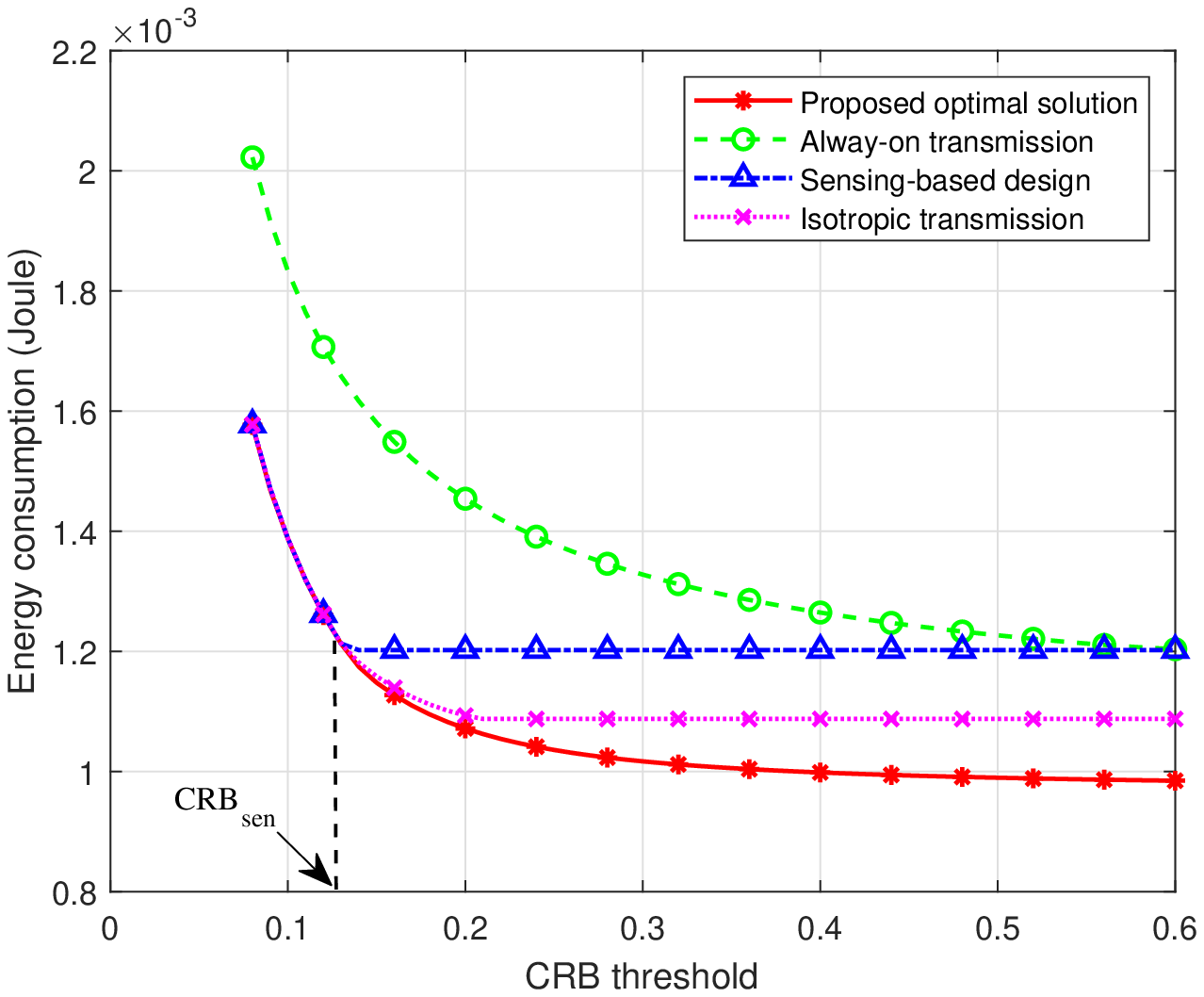}
  \caption{Energy consumption versus CRB threshold $\Gamma$ with $R=15\text{bps/Hz}$ and $d=100\text{m}$.}
  \label{CRBvery_deficient}
\end{figure}
\begin{figure}[ht]
  \centering
  \includegraphics[width=80mm]{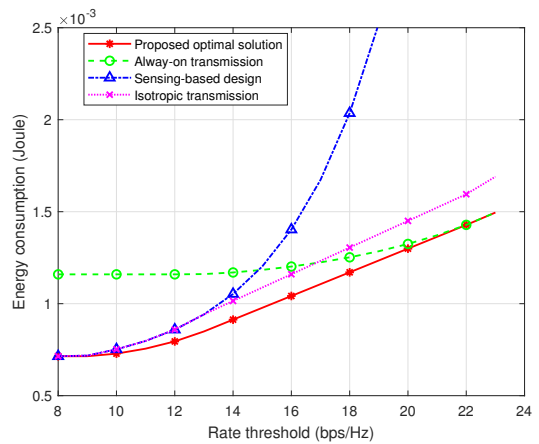}
  \caption{Energy consumption versus rate threshold $R$ with $\Gamma=0.9$ and $d=100\text{m}$.}
  \label{ratevery_deficient}
\end{figure}

Next, we consider the case when the communication channel $\boldsymbol{H}_c$ is rank-deficient, in which the number of receive antennas at CU is set as $N_c=4$, which is less than the number of transmit antennas $M=6$ at the BS. In this case, the communication-based scheme does not work, as the achieved CRB would become infinite, thus making the CRB constraint infeasible. 

Fig.~\ref{allocation} shows the power allocation behaviors with $\Gamma=0.18$ and $R=15\text{bps/Hz}$. It is observed that for the scheme with communication task only, the achieved power allocations are only allocated over the first $r=4$ subchannels, which make the correspondingly achieved CRB become infinite. Furthermore, it is observed that for the proposed optimal solution, the achieved power allocations over the first $r=4$ subchannels are monotonically non-increasing while those over the remaining $M-r=2$ channels are the same but lower than the first $r$ channels, which is consistent with Proposition \ref{Prop power allocation}. Fig.~\ref{durationrankdeficient} shows the corresponding resultant ``on'' duration for active transmission. Similar observations are made as in Fig.~\ref{durationfullrank}.

Fig.~\ref{CRBvery_deficient} and Fig.~\ref{ratevery_deficient} show the energy consumption versus the CRB threshold $\Gamma$ and the rate threshold $R$, respectively. It is shown that the proposed optimal solution  still outperforms other benchmark schemes. In addition, it is shown in Fig.~\ref{CRBvery_deficient} that when $\Gamma$ is small, the achieved energy consumption by the isotropic transmission performs close to the proposed optimal solution, which is similar as the case with full-rank $\boldsymbol{H}_c$ in Section \ref{Sec full}. However, it is observed from Fig.~\ref{ratevery_deficient} that when rate threshold $R$ is large, the proposed optimal solution significantly outperforms the isotropic transmission, which is different from the case with full-rank $\boldsymbol{H}_c$. The reason is explained as follows. For the proposed optimal solution, the first $r=4$ subchannels serve as ISAC channels for both communications and sensing, while the remaining $M-r=2$ subchannels serve as dedicated sensing channel for sensing task only. As a result, only the power allocated over the first $r=4$ subchannels needs to be increased to meet the increasing rate requirement of $R$. By contrast, the isotropic transmission needs to increase the power allocations over all subchannels to meet the rate constraint. This thus leads to the significantly increased energy consumption for the isotropic transmission, as compared to the optimal solution.

\section{Conclusion}
This paper investigated the energy efficiency of the MIMO ISAC system with one extended target and one multi-antenna CU,  by considering the practical on-off non-transmission power. In particular, we proposed the optimal transmit covariance and on-off control solution in semi-closed form to minimize the energy consumption at the BS over a finite transmission block, while ensuring the rate and CRB constraints for ISAC. The proposed design was shown to unify the energy-efficient and spectrum-efficient communications and sensing designs, and outperform other benchmark schemes, especially when both the rate and CRB constraints become stringent. How to extend the developed solutions to scenarios with other targets and communication models (e.g., the case with multiple point targets and that with multiuser communications) and those with coordinated ISAC among multiple BSs is interesting directions worthy future study. 
\appendix
\addcontentsline{toc}{chapter}{APPENDICES}

\begin{appendices}
\subsection{Proof of Proposition \ref{Optimal solution only sen}}\label{Appendix only sen}
First, it is clear that at the optimality of (P3), it follows from \cite{9652071} that the isotropic transmission is optimal for minimizing the CRB, i.e., we have $\boldsymbol{Q} =  p\boldsymbol{I}_M$, where $p$ is an optimization variable to be decided. By substituting this, problem (P3) is equivalently reformulated as
    \begin{subequations}
      \begin{align}
      (\text{P3.1}):~&{\mathop \text{min}\limits_{p \geq  0, \tau}}\quad{\frac{M}{\eta}p\tau+P_c\tau} \label{objective function in reformulated sensing-dominant}  \\
       &\quad {\text{s.t.}} \quad{\frac{\sigma_s^2N_sM}{ B\tau p}\leq \Gamma } \\
       &\quad\quad\quad\eqref{Q and tau constraint P1}\nonumber.
     \end{align} 
    \end{subequations} 
    For problem (P3.1), the objective value is monotonically increasing with respect to term $p\tau$. Therefore, to minimize the total energy consumption, the optimality is attained at $\frac{\sigma_s^2N_sM}{ B\tau p} = \Gamma$. By substituting this in \eqref{objective function in reformulated sensing-dominant}, problem (P3.1) is further reformulated as 
    \begin{subequations}
      \begin{align}
      (\text{P3.2}):~&{\mathop \text{min}\limits_{ \tau}}\quad{\frac{\sigma_s^2N_sM^2}{ \eta B\Gamma}+P_c\tau} \label{objective only w.r.t t} \\
       &~{\text{s.t.}} \quad\eqref{Q and tau constraint P1}. \nonumber
     \end{align} 
    \end{subequations} 
    It is evident that the objective is monotonically increasing with respect to $\tau$. To minimize the objective value in \eqref{objective only w.r.t t}, we have $\tau= T_{\text{min}}$. By substituting it to $\frac{\sigma_s^2N_sM}{ B\tau p} = \Gamma$, we have $p=\frac{\sigma_s^2N_sM}{BT_{\textup{min}}\Gamma}$. Therefore, the optimal solution to problem (P3) in Proposition \eqref{Optimal solution only sen} is obtained.
\subsection{Proof of Lemma \ref{lemma diagonal}} \label{Appendix diagonal}
We prove this propsoition via contradiction, by assuming that the optimal positive definite $\tilde{\boldsymbol{E}}$ is not diagonal. Then we define $\tilde{\boldsymbol{E}}_{\text{diag}}=\tilde{\boldsymbol{E}}\circ \boldsymbol{I}$ as a diagonal matrix with its corresponding diagonal elements being the same as $\tilde{\boldsymbol{E}}$. Then we have 
\begin{align}
  &\text{tr}(\tilde{\boldsymbol{E}}_{\text{diag}})=\text{tr}(\tilde{\boldsymbol{E}}), \label{lemma for diagonal trace} \\
  &\text{tr}\{(\tilde{\boldsymbol{E}}_{\text{diag}})^{-1}\}\leq\text{tr}\{(\tilde{\boldsymbol{E}})^{-1}\}, \label{lemma for diagonal trace inv} \\
  &\text{log}_2\text{det}\left(\boldsymbol{I}_{N_c}+\frac{\boldsymbol{\Sigma}^2\tilde{\boldsymbol{E}}_{\text{diag}}}{\sigma_c^2 \tau}\right)\geq\text{log}_2\text{det}\left(\boldsymbol{I}_{N_c}+\frac{\boldsymbol{\Sigma}^2\tilde{\boldsymbol{E}}}{\sigma_c^2 \tau}\right) ,\label{lemma for diagonal log}
\end{align}  
where \eqref{lemma for diagonal trace inv} follows based on \cite[Lemma 1]{1327814}, \eqref{lemma for diagonal log} holds based on the Hadamard inequality \cite{horn2012matrix}. Therefore, $\tilde{\boldsymbol{E}}_{\text{diag}}$ is also feasible for problem (P4.1) and achieves the same objective  value as $\tilde{\boldsymbol{E}}$. Notice that the inequalities hold in both \eqref{lemma for diagonal trace inv} and \eqref{lemma for diagonal log}, which imply that the BS may further reduce the energy consumption while ensuring the rate and CRB constraints. This thus contradicts the presumption that the non-diagonal $\tilde{\boldsymbol{E}}$ is optimal. As a result, it follows that the optimal solution of $\tilde{\boldsymbol{E}}$ to problem (P4.1) should be diagonal.

\subsection{Proof of Proposition \ref{Proposition of relation between ei and t}} \label{Appendix Lagrange}
First, we obtain the optimal solution of (P4.2) based on the Karush-Kuhn-Tucker conditions. For the optimal primal and dual solutions $(\left\{e_i^*\right\},  \tau^*, \gamma^*, \nu^*)$, it follows that
\begin{subequations}
\begin{align}
&\gamma^*\left(\frac{\sigma_s^2 N_s}{B} \sum_{i=1}^M \frac{1}{e_i^*}-\Gamma\right)=0, \label{complementary 1}\\
&\nu^*\left(\frac{\tau^*}{T_{\text{max}}} \sum_{i=1}^r \log _2\left(1+\frac{1}{\sigma_c^2 \tau^*} \lambda_i{ }^2 e^*_i\right)-R\right)=0, \label{complementary 2}\\
&\frac{1}{\eta}-\frac{\gamma^*\sigma_s^2N_s}{B}\frac{1}{(e^*_i)^2}-\frac{\nu^*}{\text{ln}2 T_{\text{max}}}\left(\frac{\frac{\lambda_i^2}{\sigma_c^2}}{1+\frac{\lambda_i^2e^*_i}{\sigma_c^2\tau^*}}\right)=0,1\leq i \leq r,  \label{derivatives of e1 within r}\\
&\frac{1}{\eta}-\frac{\gamma^*\sigma_s^2N_s}{B}\frac{1}{(e^*_i)^2}=0,~r+1\leq i \leq M,  \label{derivatives of e1 r+1 to M}\\
&P_c-\frac{\nu^*}{T_{\text{max}}}\left(\sum_{i=1}^r\text{log}_2(1+\frac{\lambda_i^2e^*_i}{\sigma_c^2\tau^*})+\frac{1}{\text{ln}2}\sum_{i=1}^r(\frac{1}{1+\frac{\lambda_i^2e^*_i}{\sigma_c^2\tau^*}})\right. \nonumber\\
&\quad\left.-\frac{r}{\text{ln}2}\right)=0, \label{derivatives of L} 
\end{align}
\end{subequations}
where \eqref{complementary 1} and \eqref{complementary 2} denote the complementary slackness condition, the left-hand side of \eqref{derivatives of e1 within r} and \eqref{derivatives of e1 r+1 to M} are the partial derivatives of $\mathcal{L} (\boldsymbol{e},  \tau, \gamma, \nu)$ in \eqref{Lagrangian for primal problem} with respect to $e_i$ for $i \in \{1,\ldots,r\}$ and $i \in \{r+1,\ldotp,M\}$, respectively, and \eqref{derivatives of L} is the partial derivative of $\mathcal{L} (\boldsymbol{e},  \tau, \gamma, \nu)$ with respect to $\tau$. Therefore, based on \eqref{derivatives of e1 within r}, we have the cubic equation as 
\begin{align}
  \frac{{\lambda _i^2}}{\eta {\sigma _c^2}\tau^*}(e^*_i)^3+( \frac{1}{\eta } - &\frac{{{\nu^*}}}{{\text{ln} 2T_{\text{max}}}}\frac{{\lambda _i^2}}{{\sigma _c^2}})(e^*_i)^2 - \frac{\gamma^*\sigma_s^2N_s}{B}\frac{{\lambda _i^2}}{{\sigma _c^2}\tau^*}e^*_i  \nonumber\\
  &- \frac{\gamma^*\sigma_s^2N_s}{B} =0,~i \in \{1,\ldots,r\}. \label{cubic ellipsoid}
\end{align} According to the Cardano's formula for solving a cubic equation, we have \eqref{cubic equation}.
Based on \eqref{derivatives of e1 r+1 to M}, we have 
\begin{align} \label{ei ellipsoid}
  e^*_{i}=\sqrt {{\frac{\eta\gamma^*\sigma_s^2N_s}{B}}}, ~\forall i\in \{r+1,\ldots,M\}.
\end{align}
Furthermore, based on \eqref{derivatives of L}, we have \eqref{bisection equation}.

Next, it remains to solve the dual problem (D4.2) to find the optimal dual variables $(\gamma^*, \nu^* )$. As (D4.2) is convex but non-differentiable in general, we use subgradient based methods such as ellipsoid method. For the objective function $\mathcal{G}(\gamma, \nu)$ in problem (D4.2), the subgradient for $\left\{\gamma,\nu \right\}$ is given by 
  \begin{align}
    &\left[\frac{\sigma _s^2{N_s}}{B}\sum\limits_{i = 1}^M {\frac{1}{{{e^*_i}}}}  - {\Gamma},{R} - \frac{\tau^* }{T_{\text{max}}}\sum\limits_{i = 1}^r {\text{log}_2\left( {1 + \frac{{\lambda _i}^2{e^*_i}}{{\sigma _c^2\tau^*}}} \right)}\right]^{T}.
  \end{align}
For the constraints $\gamma \geq0$ and $ \nu  \geq0$, the corresponding subgradients are given as $\left[1, 0\right]^{T}$ and $\left[0, 1\right]^{T}$, respectively. Accordingly, we obtain the optimal solution $(\gamma^*,\nu^*)$ to ($\text{D4.2}$). 

By substitute $(\gamma^*,\nu^*)$ back to \eqref{cubic ellipsoid} and \eqref{ei ellipsoid}, we obtain \eqref{cubic equation} and \eqref{ei r1}, respectively. This thus completes the proof of Propsootion \ref{Proposition of relation between ei and t}.

\subsection{Proof of Proposition \ref{Prop power allocation}}\label{Appendix power allocation}
We present the proof by considering the case with $r < M$, while that with $r=M$ can be verified similarly and thus is omitted. First, based on \eqref{ei r1}, we easily obtain $e_{r+1}^*=\ldots=e_{M}^*>0$. Next, we prove $e_{r}^*\geq e_{r+1}^*$ by contradiction. By assuming $e_{r}^*<e_{r+1}^*=\sqrt {\frac{\eta{{\gamma ^*}\sigma _s^2{N_s}}}{B}}$, we have $\frac{1}{\eta}-\frac{\gamma^*\sigma_s^2N_s}{B}\frac{1}{(e_r^*)^2}<0$. Then by substituting this inequality into \eqref{power within r}, we have 
\begin{align}
  \frac{1}{\eta}-\frac{\gamma^*\sigma_s^2N_s}{B}\frac{1}{(e_r^*)^2}=\frac{\nu^*}{\text{ln}2T_{\text{max}}}\left(\frac{\frac{\lambda_r^2}{\sigma_c^2}}{1+\frac{\lambda_r^2e_r^*}{\sigma_c^2\tau^*}}\right)<0,
\end{align}
where there is a contradiction since dual variables and other parameters are non-negative. Accordingly, we have $e_{r}^* \geq e_{r+1}^*$. Finally, in remains to prove $e_1^*\geq e_2^*\geq \ldots \geq e_r^*$. Similarly, we use contradiction again by assuming $e_{i}^*<e_{i+1}^*,~i\leq r-1$, then 
\begin{align} \label{proof of rem 1}
  &\frac{1}{\eta}-\frac{\gamma^*\sigma_s^2N_s}{B}\frac{1}{(e_i^*)^2}=\frac{\nu^*}{\text{ln}2T_{\text{max}}}\left(\frac{\frac{\lambda_i^2}{\sigma_c^2}}{1+\frac{\lambda_i^2e_i^*}{\sigma_c^2\tau^*}}\right)\nonumber \\
  &\quad\quad<\frac{1}{\eta}-\frac{\gamma^*\sigma_s^2N_s}{B}\frac{1}{(e_{i+1}^*)^2}=\frac{\nu^*}{\text{ln}2T_{\text{max}}}\left(\frac{\frac{\lambda_{i+1}^2}{\sigma_c^2}}{1+\frac{\lambda_{i+1}^2e_{i+1}^*}{\sigma_c^2\tau^*}}\right).
\end{align}
Recall that we have $\lambda_1\geq\ldots \geq \lambda_r$ for the singular values of communication channel $\boldsymbol{H}_c$. It is easy to obtain 
\begin{align}
  \frac{\nu^*}{\text{ln}2T_{\text{max}}}\left(\frac{\frac{\lambda_i^2}{\sigma_c^2}}{1+\frac{\lambda_i^2e_i^*}{\sigma_c^2\tau^*}}\right)>\frac{\nu^*}{\text{ln}2T_{\text{max}}}\left(\frac{\frac{\lambda_{i+1}^2}{\sigma_c^2}}{1+\frac{\lambda_{i+1}^2e_{i+1}^*}{\sigma_c^2\tau^*}}\right),
\end{align}
which is contradictory to \eqref{proof of rem 1}. As a result, $e_1^*\geq e_2^*\geq \ldots \geq e_r^*$ can be proved. Therefore, it follows that $e_1^*\geq e_2^*\geq \ldots \geq e_r^*\geq e_{r+1}^*= \ldots =e_M^* >0$. As we define $p_i^*=\frac{e_i^*}{\tau^*}$, such that we have $p_1^*\geq p_2^*\geq \ldots \geq p_r^*\geq p_{r+1}^*= \ldots =p_M^* >0$. This thus completes the proof of Proposition \ref{Prop power allocation}.
\end{appendices}

\bibliographystyle{IEEEtran}

\bibliography{IEEEabrv,reference}

\begin{thebibliography}{10}
\providecommand{\url}[1]{#1}
\csname url@samestyle\endcsname
\providecommand{\newblock}{\relax}
\providecommand{\bibinfo}[2]{#2}
\providecommand{\BIBentrySTDinterwordspacing}{\spaceskip=0pt\relax}
\providecommand{\BIBentryALTinterwordstretchfactor}{4}
\providecommand{\BIBentryALTinterwordspacing}{\spaceskip=\fontdimen2\font plus
\BIBentryALTinterwordstretchfactor\fontdimen3\font minus
  \fontdimen4\font\relax}
\providecommand{\BIBforeignlanguage}[2]{{%
\expandafter\ifx\csname l@#1\endcsname\relax
\typeout{** WARNING: IEEEtran.bst: No hyphenation pattern has been}%
\typeout{** loaded for the language `#1'. Using the pattern for}%
\typeout{** the default language instead.}%
\else
\language=\csname l@#1\endcsname
\fi
#2}}
\providecommand{\BIBdecl}{\relax}
\BIBdecl

\bibitem{9737357}
F.~Liu, Y.~Cui, C.~Masouros, J.~Xu, T.~X. Han, Y.~C. Eldar, and S.~Buzzi,
  ``Integrated sensing and communications: Toward dual-functional wireless
  networks for {6G} and beyond,'' \emph{IEEE J. Sel. Areas Commun.}, vol.~40,
  no.~6, pp. 1728--1767, Jun. 2022.

\bibitem{8972666}
Z.~Feng, Z.~Fang, Z.~Wei, X.~Chen, Z.~Quan, and D.~Ji, ``Joint radar and
  communication: A survey,'' \emph{China Commun.}, vol.~17, no.~1, pp. 1--27,
  Jan. 2020.

\bibitem{9652071}
F.~Liu, Y.-F. Liu, A.~Li, C.~Masouros, and Y.~C. Eldar, ``Cram{\'e}r-{R}ao
  bound optimization for joint radar-communication beamforming,'' \emph{IEEE
  Trans. Signal Process.}, vol.~70, pp. 240--253, Dec. 2021.

\bibitem{8999605}
F.~Liu, C.~Masouros, A.~P. Petropulu, H.~Griffiths, and L.~Hanzo, ``Joint radar
  and communication design: Applications, state-of-the-art, and the road
  ahead,'' \emph{IEEE Trans. Commun.}, vol.~68, no.~6, pp. 3834--3862, Jun.
  2020.

\bibitem{9842350}
Y.~Huang, Y.~Fang, X.~Li, and J.~Xu, ``Coordinated power control for network
  integrated sensing and communication,'' \emph{IEEE Trans. Veh. Technol.},
  vol.~71, no.~12, pp. 13\,361--13\,365, Dec. 2022.

\bibitem{9606831}
Y.~Cui, F.~Liu, X.~Jing, and J.~Mu, ``Integrating sensing and communications
  for ubiquitous {IoT}: Applications, trends, and challenges,'' \emph{IEEE
  Netw.}, vol.~35, no.~5, pp. 158--167, Sep. 2021.

\bibitem{9916163}
Z.~Lyu, G.~Zhu, and J.~Xu, ``Joint maneuver and beamforming design for
  {UAV}-enabled integrated sensing and communication,'' \emph{IEEE Trans.
  Wireless Commun.}, vol.~22, no.~4, pp. 2424--2440, Apr. 2023.

\bibitem{9127852}
D.~Ma, N.~Shlezinger, T.~Huang, Y.~Liu, and Y.~C. Eldar, ``Joint
  radar-communication strategies for autonomous vehicles: Combining two key
  automotive technologies,'' \emph{IEEE Signal Process. Mag.}, vol.~37, no.~4,
  pp. 85--97, Jul. 2020.

\bibitem{telatar1999capacity}
E.~Telatar, ``Capacity of multi-antenna {Gaussian} channels,'' \emph{Eur.
  Trans. Telecommun.}, vol.~10, no.~6, pp. 585--595, 1999.

\bibitem{li2007mimo}
J.~Li and P.~Stoica, ``{{MIMO} }radar with colocated antennas,'' \emph{IEEE
  Signal Process Mag.}, vol.~24, no.~5, pp. 106--114, Sep. 2007.

\bibitem{haimovich2007mimo}
A.~M. Haimovich, R.~S. Blum, and L.~J. Cimini, ``{MIMO} radar with widely
  separated antennas,'' \emph{IEEE Signal Process Mag.}, vol.~25, no.~1, pp.
  116--129, Dec. 2007.

\bibitem{hua2022mimo}
H.~Hua, T.~X. Han, and J.~Xu, ``{MIMO} integrated sensing and communication:
  {CRB}-rate tradeoff,'' \emph{arXiv:2209.12721}, 2022.

\bibitem{8386661}
F.~Liu, L.~Zhou, C.~Masouros, A.~Li, W.~Luo, and A.~Petropulu, ``Toward
  dual-functional radar-communication systems: Optimal waveform design,''
  \emph{IEEE Trans. Signal Process.}, vol.~66, no.~16, pp. 4264--4279, Aug.
  2018.

\bibitem{hua2021optimal}
H.~Hua, J.~Xu, and T.~X. Han, ``Optimal transmit beamforming for integrated
  sensing and communication,'' \emph{IEEE Trans. Veh. Technol}, pp. 1--16,
  early access, Mar. 2023, doi: {\color{blue}{10.1109/TVT.2023.3262513}}.

\bibitem{9124713}
X.~Liu, T.~Huang, N.~Shlezinger, Y.~Liu, J.~Zhou, and Y.~C. Eldar, ``Joint
  transmit beamforming for multiuser {MIMO} communications and {MIMO} radar,''
  \emph{IEEE Trans. Signal Process.}, vol.~68, pp. 3929--3944, Jun. 2020.

\bibitem{9531484}
C.~Xu, B.~Clerckx, S.~Chen, Y.~Mao, and J.~Zhang, ``Rate-splitting multiple
  access for multi-antenna joint radar and communications,'' \emph{IEEE J. Sel.
  Topics Signal Process.}, vol.~15, no.~6, pp. 1332--1347, Nov. 2021.

\bibitem{8922617}
T.~Huang, W.~Yang, J.~Wu, J.~Ma, X.~Zhang, and D.~Zhang, ``A survey on green
  {6G} network: Architecture and technologies,'' \emph{IEEE Access}, vol.~7,
  pp. 175\,758--175\,768, Dec. 2019.

\bibitem{6082514}
Z.~Chong and E.~Jorswieck, ``Energy-efficient power control for {MIMO}
  time-varying channels,'' in \emph{Proc. IEEE Online Conf. Green Commun.
  (GreenCom)}, 2011, pp. 92--97.

\bibitem{5683485}
R.~S. Prabhu and B.~Daneshrad, ``Energy-efficient power loading for a
  {MIMO-SVD} system and its performance in flat fading,'' in \emph{Proc. IEEE
  Global Telecommun. Conf. (GLOBECOM)}, Dec. 2010, pp. 1--5.

\bibitem{6409501}
J.~Xu and L.~Qiu, ``Energy efficiency optimization for {MIMO} broadcast
  channels,'' \emph{IEEE Trans. Wireless Commun.}, vol.~12, no.~2, pp.
  690--701, Feb. 2013.

\bibitem{6514948}
J.~Xu and R.~Zhang, ``Throughput optimal policies for energy harvesting
  wireless transmitters with non-ideal circuit power,'' \emph{IEEE J. Sel.
  Areas Commun.}, vol.~32, no.~2, pp. 322--332, Feb. 2014.

\bibitem{kaushik2021hardware}
A.~Kaushik, C.~Masouros, and F.~Liu, ``Hardware efficient joint
  radar-communications with hybrid precoding and {RF} chain optimization,'' in
  \emph{Proc. IEEE Int. Conf. Commun. (ICC)}, Jun. 2021, pp. 1--6.

\bibitem{kaushik2022green}
A.~Kaushik, E.~Vlachos, C.~Masouros, C.~Tsinos, and J.~Thompson, ``Green joint
  radar-communications: {RF} selection with low resolution {DACs} and hybrid
  precoding,'' in \emph{Proc. IEEE Int. Conf. Commun. (ICC)}, May 2022, pp.
  1--6.

\bibitem{he2022energy}
Z.~He, W.~Xu, H.~Shen, Y.~Huang, and H.~Xiao, ``Energy efficient beamforming
  optimization for integrated sensing and communication,'' \emph{IEEE Wireless
  Commun. Lett.}, vol.~11, no.~7, pp. 1374--1378, Jul. 2022.

\bibitem{9797869}
O.~Dizdar, A.~Kaushik, B.~Clerckx, and C.~Masouros, ``Energy efficient
  dual-functional radar-communication: Rate-splitting multiple access,
  low-resolution {DACs}, and {RF} chain selection,'' \emph{IEEE Open J. Commun.
  Soc.}, vol.~3, pp. 986--1006, Jun. 2022.

\bibitem{li1996adaptive}
J.~Li and P.~Stoica, ``{An adaptive filtering approach to spectral estimation
  and SAR imaging},'' \emph{IEEE Trans. Signal Process.}, vol.~44, no.~6, pp.
  1469--1484, Jun. 1996.

\bibitem{schmidt1986multiple}
R.~Schmidt, ``{Multiple emitter location and signal parameter estimation},''
  \emph{IEEE Trans. Antennas Propag.}, vol.~34, no.~3, pp. 276--280, Mar. 1986.

\bibitem{kay1993fundamentals}
S.~M. Kay, \emph{{Fundamentals of Statistical Signal Processing: Estimation
  Theory}}.\hskip 1em plus 0.5em minus 0.4em\relax Prentice-Hall, Inc., 1993.

\bibitem{4276989}
P.~Stoica, J.~Li, and Y.~Xie, ``On probing signal design for {MIMO} radar,''
  \emph{IEEE Trans. Signal Process.}, vol.~55, no.~8, pp. 4151--4161, Jul.
  2007.

\bibitem{4838872}
Z.~Ben-Haim and Y.~C. Eldar, ``On the constrained {Cram{\'e}r-{R}ao} bound with
  a singular fisher information matrix,'' \emph{IEEE Signal Process. Lett.},
  vol.~16, no.~6, pp. 453--456, Jun. 2009.

\bibitem{richards2014fundamentals}
M.~A. Richards, \emph{Fundamentals of {Radar Signal Processing}}.\hskip 1em
  plus 0.5em minus 0.4em\relax New York, NY, USA: McGraw-Hill Educ., 2014.

\bibitem{1327814}
S.~Ohno and G.~Giannakis, ``Capacity maximizing {MMSE}-optimal pilots for
  wireless {OFDM} over frequency-selective block {Rayleigh}-fading channels,''
  \emph{IEEE Trans. Inf. Theory}, vol.~50, no.~9, pp. 2138--2145, Aug. 2004.

\bibitem{horn2012matrix}
R.~A. Horn and C.~R. Johnson, \emph{Matrix analysis}.\hskip 1em plus 0.5em
  minus 0.4em\relax Cambridge university press, 2012.

\end{thebibliography}
\end{document}